\documentclass[pre,showpacs,showkeys,preprintnumbers,amsmath,amssymb,superscriptaddress,twocolumn]{revtex4-2}
\usepackage{graphicx}
\usepackage{amsfonts}
\usepackage{url}
\usepackage{epstopdf}
\usepackage{color}
\usepackage{bm}
\usepackage[colorlinks=true,linkcolor=blue,citecolor=red]{hyperref}%
\usepackage{comment}

\def\qa{q_{\rm a}}
\def\qc{q_{\rm c}}
\def\qccrit{{q_{\rm c}^{\rm crit}}}

\def\Gammaa{\Gamma_{\rm a}}
\def\Gammac{\Gamma_{\rm c}}
\def\Gammar{\Gamma_{\rm r}}

\def\A{{\mathcal A}}

\def\g{\mathfrak{g}}
\def\G{{\mathcal{G}}}

\def\a{\mathrm{a}}
\def\c{\mathrm{c}}

\def\L{\mathcal L}
\def\N{\mathcal N}

\def\ve{\varepsilon}

\def\pa{\partial\Omega}
\def\E{{\mathbb E}}
\def\P{{\mathbb P}}
\def\R{{\mathbb R}}
\def\T{{\mathcal T}}

\def\x{\bm{x}}
\def\X{\bm{X}}
\def\y{\bm{y}}

\begin{document}

\title{Population dynamics of surface-mediated autocatalytic processes}

\author{Denis~S.~Grebenkov}
 \email{denis.grebenkov@polytechnique.edu}
\affiliation{
Laboratoire de Physique de la Mati\`{e}re Condens\'{e}e, \\ 
CNRS -- Ecole Polytechnique, Institut Polytechnique de Paris, 91120 Palaiseau, France}

\author{Yilin~Ye}
 \email{yilin.ye@polytechnique.edu}
\affiliation{
Laboratoire de Physique de la Mati\`{e}re Condens\'{e}e, \\ 
CNRS -- Ecole Polytechnique, Institut Polytechnique de Paris, 91120 Palaiseau, France}

\date{\today}

\begin{abstract}
We investigate the population dynamics of surface-mediated
autocatalytic processes, in which particles diffuse in a complex
environment towards surface regions where they can be either killed or
replicated.  These opposite mechanisms compete with each other and
lead to a sophisticated stochastic evolution of the population size.
We provide a systematic analysis of the generating function of the
population size.  We also deduce its distribution, mean, variance and
higher-order moments.  For this purpose, we employ several equivalent
descriptions of these quantities in terms of nonlinear integral
equations and partial differential equations with nonlinear boundary
conditions.  We inspect the long-time behavior of the population
dynamics in three regimes when the mean population size vanishes,
reaches a steady-state level, or grows exponentially.  A numerical
solution of the underlying integral equations and independent Monte
Carlo simulations support our theoretical predictions.
\end{abstract}


%02.50.-r       (Probability theory, stochastic processes, and statistics)
%05.40.-a 	Fluctuation phenomena, random processes, noise, and Brownian motion
%02.70.Rr       (General statistical methods)
%05.10.Gg 	Stochastic analysis methods (Fokker-Planck, Langevin, etc.) 

%02.50.Ey 	Stochastic processes  (Probability theory, stochastic processes, and statistics)


\maketitle

\section{Introduction}

Diffusion-controlled reactions are ubiquitous in physical, chemical
and biological phenomena
\cite{North66,Wilemski73,Calef83,Berg85,Rice85,Grebenkov07,Bressloff13,Holcman13,Benichou14,Galanti16}.
In a typical setting, a particle diffuses in a confinement until
hitting a reactive region, on which it may relax its excited state, be
trapped, chemically transformed, expelled from the domain, or killed
\cite{Grebenkov23g}.  Irrespective of the microscopic origin of
the surface reaction mechanism, a common mathematical description
employs the Robin boundary condition
\begin{equation} \label{eq:RobinC}
-\partial_n C = q C,
\end{equation}
in which the diffusive flux of particles from the bulk onto the
boundary on the left is proportional to their concentration $C$ on
that boundary \cite{Collins49} (see \cite{Grebenkov20,Piazza22} for
further discussions).  The proportionality coefficient $q$ quantifies
the rate of the surface reaction and ranges from $0$ for inert
boundary (no reaction) to $+\infty$ for a perfect sink (instant
reaction upon the first arrival onto the reactive region).  Since the
particles are progressively sunk from the system through the reactive
region, their concentration declines and vanishes with time.  Numerous
former studies inspected how the absorption rate $q$ affects the
production rate of a chemical reactor, functioning of living cells and
respiration organs, and efficiency of diffusion-mediated search
processes (see
\cite{Redner,Ben-Avraham,Krapivsky,Schuss,Metzler,Lindenberg,Grebenkov}
and references therein).

Quite naturally, one may wonder what happens for a {\it negative}
parameter $q$?  In this case, the right-hand side of the Robin
boundary condition (\ref{eq:RobinC}) is negative, meaning that the
direction of the diffusive flux is reversed and is now oriented from
the boundary into the bulk.  In other words, the region with negative
$q$ produces particles on the boundary and injects them into the bulk.
However, as this production is proportional to the concentration $C$,
the mechanism is different from a well-studied injection of particles
into the bulk with a prescribed rate.  Such a negative reactivity can
be rationalized by a microscopic probabilistic model of
boundary-catalytic branching processes or, equivalently,
surface-mediated autocatalytic reactions \cite{Grebenkov26}.  At each
arrival onto the catalytic region, a particle can split with the rate
$|q|$ into two particles that continue diffusing independently from
each other.  The mean population size of this system was shown to obey
the Robin boundary condition with the negative parameter $q$
\cite{DelGrosso76,Grebenkov26}.  Since the number of particles (and
thus the number of degrees of freedom in the system) changes randomly,
a full description of this diffusion-reaction dynamics requires
elaborate mathematical tools such as measure-valued stochastic
processes (or superprocesses), random snakes,
etc. \cite{LeGall,Dynkin,Dawson99,Klenke00,Kesten03,Englander04,Delmas05,Morters05,Englander07,Bocharov14,Bulinskaya18}.
However, the random number of particles at time $t$ that we call the
population size $\N(t)$, can be fully described by using conventional
probabilistic tools.  In fact, two equivalent descriptions were
established in \cite{Grebenkov26a} for a general diffusion process: a
nonlinear integral equation and a backward Fokker-Planck equation with
nonlinear boundary conditions.

In this paper, we provide a systematic analysis of surface-mediated
autocatalytic processes for ordinary diffusion in a bounded domain.
Section \ref{sec:model} recalls the microscopic model introduced in
\cite{Grebenkov26a} and its probabilistic description.  In
Sec. \ref{sec:Gs}, we present different ways to obtain the generating
function of the population size: a renewal-type integral equation
(Sec. \ref{sec:integral}), a partial differential equation (PDE)
reformulation (Sec. \ref{sec:PDE}), and two dual representations
(Sec. \ref{sec:dual}).  We also deduce integral and differential
equations for the distribution of the population size
(Sec. \ref{sec:distrib}).  Section \ref{sec:Gs_asympt} presents our
main results on the population dynamics at long times.  In particular,
we identify three asymptotic regimes and discuss the long-time
behavior of the generating function and of the distribution.  In turn,
the long-time behavior of the moments of the population size is
analyzed in Sec. \ref{sec:moments}.  Section
\ref{sec:discussion} concludes the paper by summarizing the main
results and identifying open problems and future perspectives.  Many
technical points are relegated to Appendices, including our numerical
scheme for solving nonlinear integral equations and Monte Carlo
simulations.

\section{Model}
\label{sec:model}

\begin{figure}
\begin{center}
\includegraphics[width=0.65\columnwidth]{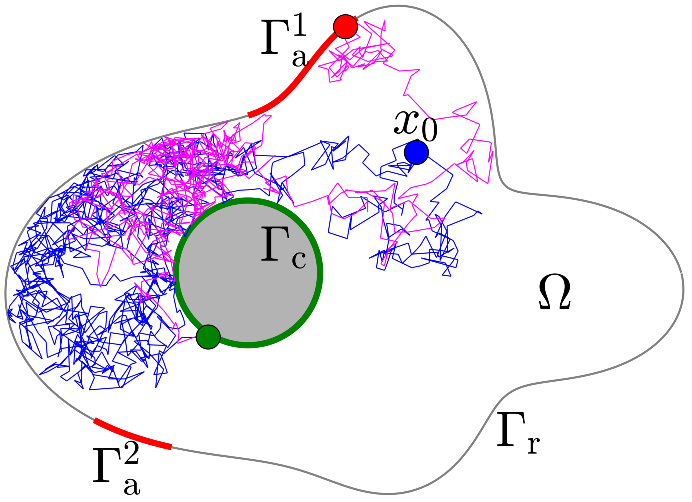} % {scheme1.eps}
\end{center}
\caption{
A schematic view of a confining domain $\Omega$, whose boundary $\pa$
is partitioned into three subsets: an absorbing region $\Gammaa$ (in
red) that can destroy each arrived particle with an absorption rate
$\qa$ via $A \to \emptyset$; an inert region $\Gammar$ (in gray) that
reflects particles back into $\Omega$; and a catalytic region
$\Gammac$ that can replicate each arrived particle into two
independent copies with a catalytic rate $\qc$ via $A\to 2A$.  Each of
three subsets can be composed of multiple disconnected pieces (e.g.,
$\Gammaa = \Gammaa^1 \cup \Gammaa^2$).  Two simulated trajectories are
shown in blue and magenta: the first particle is released from a point
$\x_0$ (blue dot) and diffuses until its binary branching (green dot)
on $\Gammac$; after this branching event, two newborn particles keep
diffusing, and one of them later disappears on $\Gammaa$ (red dot).}
\label{fig:scheme}
% load('traj.mat');
% A_Yilin4_scheme2(X,Y);
%
% A_Yilin4_scheme3;
\end{figure}

We consider a bounded connected domain $\Omega \subset\R^d$ with a
smooth boundary $\pa$, which is split into up to three disjoint
subsets (Fig. \ref{fig:scheme}): a catalytic region $\Gammac$, an
absorbing region $\Gammaa$, and the remaining reflecting region
$\Gammar = \pa \backslash (\Gammaa \cup \Gammac)$ (note that some of
these regions can be empty).  A single particle is released from a
point $\x_0 \in \overline{\Omega}$ at time $0$ and diffuses in
$\Omega$ with a constant diffusivity $D > 0$.  When the particle hits
the absorbing region $\Gammaa$, it may disappear with a small
probability proportional to a constant reactivity $\kappa$
\cite{Grebenkov03,Grebenkov06,Erban07,Singer08}.  Since absorptions
occur on the boundary, the probability of the absorption event is
controlled by the {\it boundary local time} $\ell_{t,\rm a}$, which
characterizes the time spent by the particle in a close vicinity of
$\Gammaa$ up to time $t$ (despite its name, $\ell_{t,\rm a}$ has units
of length, see Appendix \ref{sec:Amodel} for a more detailed
description).  In this light, even though the ratio $\qa = \kappa/D$
has units of inverse length, one can interpret $\qa$ as the surface
absorption ``rate'' with respect to the boundary local time.
Similarly, at each arrival onto the catalytic region $\Gammac$, the
particle may branch (or split) into two identical copies of itself
with a small probability, which is proportional to the surface
branching rate $\qc \geq 0$.  Two newborn particles are released from
the location of the branching event and diffuse independently.  Each
of these particles will disappear or branch at a later time, and so
on.

When there is no catalytic branching ($\qc = 0$), the initially
released single particle will eventually disappear.  This is the
conventional setting of diffusion-controlled reactions, which is
commonly described by looking at the survival probability of the
particle and the resulting distribution of its first-reaction time
\cite{Redner,Krapivsky,Schuss,Metzler,Lindenberg,Grebenkov,Bray13,Levernier19}.
Even if many independent particles were released at time $0$, their
probabilistic description would essentially remain a single-particle
problem.

In stark contrast, the probabilistic description of boundary-catalytic
branching processes with $\qc > 0$ is notoriously more difficult.  In
fact, even though the particles diffuse, disappear and branch
independently, a state of the system includes many degrees of freedom
(the positions and boundary local times on $\Gammaa$ and $\Gammac$ of
all particles) and, most importantly, their number changes randomly
upon branching and absorption events.  In the following, we focus on
the population dynamics and investigate how the number of particles, a
discrete-valued stochastic process $\N(t)$, evolves with time.  Even
though all other degrees of freedom can be averaged out, the branching
events lead to nonlinear boundary conditions and thus present a richer
and much more difficult problem than the conventional setting of
diffusion-controlled reactions without catalytic branching.

In order to motivate the following theoretical analysis, we start by
presenting numerical results from Monte Carlo simulations of the
population size $\N(t)$ (see Appendix \ref{sec:MC} for details).  As a
basic example, we consider diffusion inside a circular annulus $\Omega
= \{\x\in\R^2 ~:~ R < |\x| < L\}$ with the catalytic region $\Gammac$
on the inner circle of radius $R = 0.1$ and the perfectly absorbing
region $\Gammaa$ (with $\qa = \infty$) on the outer circle of radius
$L = 1$.  This example will be used throughout the manuscript for
illustrative purposes.  For each simulation, the starting point is
chosen randomly with a uniform distribution in $\Omega$.

\begin{figure}
\begin{center}
\includegraphics[width=0.99\columnwidth]{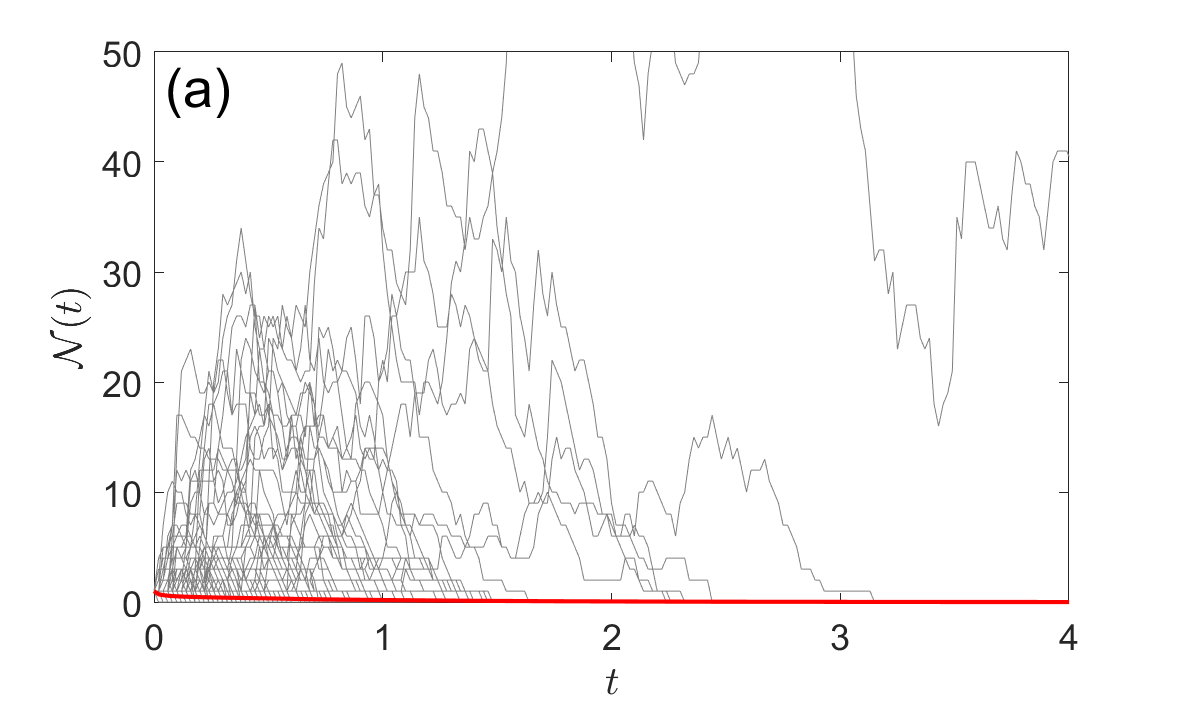} % {Nt_random_qc09.eps}
\includegraphics[width=0.99\columnwidth]{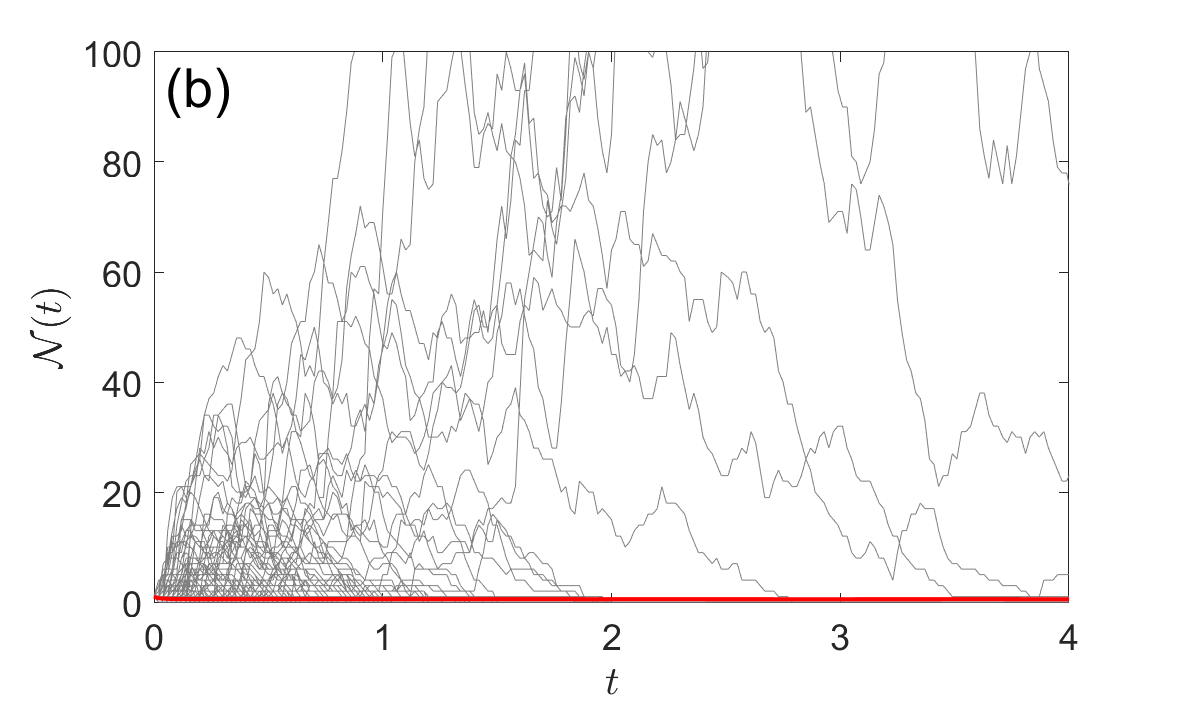} % {Nt_random_qc10.eps}
\includegraphics[width=0.99\columnwidth]{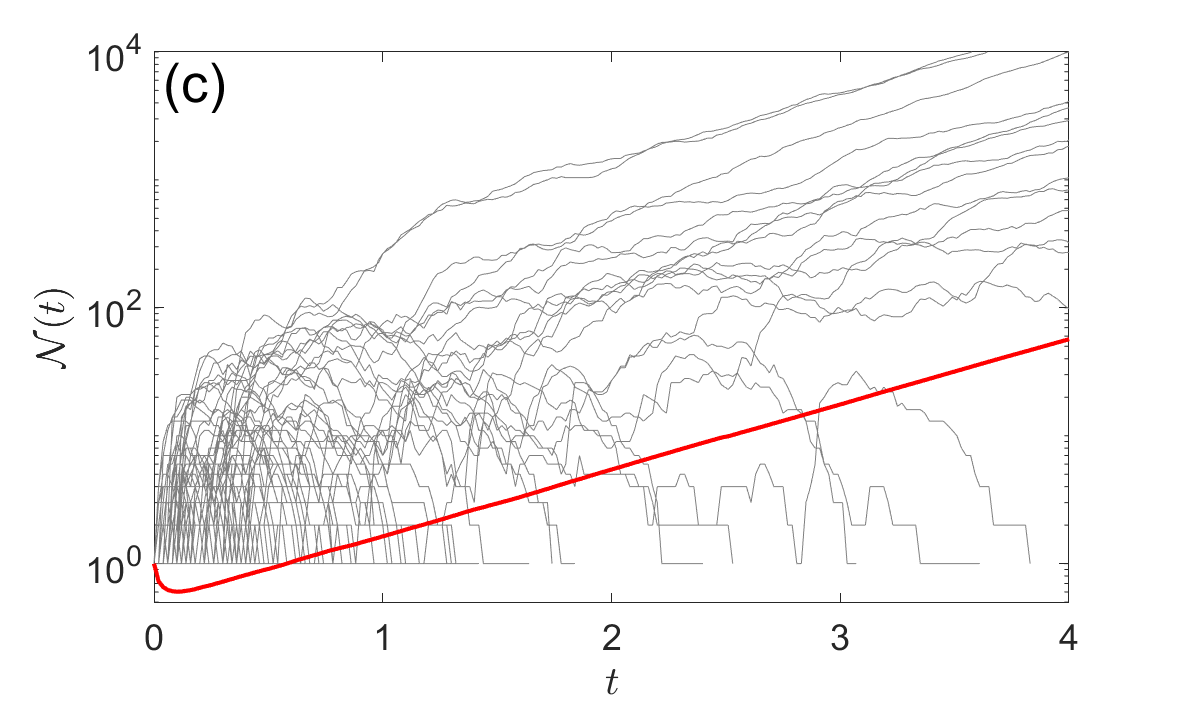} % {Nt_random_qc11.eps}
\end{center}
\caption{
1000 random realizations (gray curves) of the population dynamics in
the circular annulus with $R = 0.1$, $L = 1$, $D = 1$, $\qa = \infty$,
uniform starting point in the bulk, and three values of $\qc$: $3.91$
{\bf (a)}, $4.34$ {\bf (b)}, and $4.78$ {\bf (c)}.  Solid red line
shows the empirical mean (obtained over $10^5$ realizations, see
Appendix \ref{sec:MC}).  In the last panel, the logarithmic scale of
the vertical axis results is cutting zero values.}
\label{fig:Nt_traj}
% [Nt] = A_Yilin4_simu11(2);
% [Nt] = A_Yilin4_simu11(1);
% [Nt] = A_Yilin4_simu11(3);
\end{figure}

Figure \ref{fig:Nt_traj} presents 1000 random realizations of the
stochastic process $\N(t)$ for three values of the catalytic rate
$\qc$: $3.91$ (panel a), $4.34$ (panel b), and $4.78$ (panel c).  A
visual inspection of this figure reveals several remarkable features
of this diffusion-reaction dynamics.  First, one can observe gigantic
fluctuations between individual realizations in all three panels.  In
some realizations, the population size achieves values that are two or
three orders of magnitude larger than the mean value shown by red
curve.  Why does the mean population size not seem to be
representative?  Second, a minor variation of the catalytic rate $\qc$
by only $10\%$ among three panels dramatically changes the overall
trends in the population size.  In fact, one can see that most
realizations rapidly decline on panel (a), generally but slowly
decrease on panel (b), and exponentially growth on panel (c), in which
the logarithmic scale is used on the vertical axis.  How can one
distinguish these regimes for an arbitrary geometric setting with
given catalytic and absorption rates?  Third, the branching events can
be dominant and lead to rapidly growing populations even though the
absorbing region has an infinite absorption rate $\qa$ and is 10 times
larger than the catalytic one.  Why do the similar probabilistic
constructions of absorption and branching events result in such
dramatically different outcomes?

In the following, we answer these and many other questions by
elaborating the theoretical framework that was introduced in
\cite{Grebenkov26a} to characterize the distribution of the population
size and its moments.

\section{Generating function}
\label{sec:Gs}

The central quantity of our study is the probability generating
function
\begin{equation}
G_s(t|\x_0) = \E_{\x_0}\{ s^{\N(t)} \} \qquad (0 \leq s \leq 1)
\end{equation}
of the population size $\N(t)$, initiated by a single particle started
from $\x_0$ at time $0$ (i.e., $\N(0) = 1$).  Here the expectation
$\E$ is taken with respect to all degrees of freedom in the system,
and the subscripts highlights the starting point $\x_0$.  Rewriting
the above expectation as
\begin{equation}  \label{eq:Gs_def}
G_s(t|\x_0) = \sum\limits_{k=0}^\infty s^k \, Q_k(t|\x_0) ,
\end{equation}
one can determine the probability $Q_k(t|\x_0)$ to have $k$ particles
at time $t$:
\begin{equation}  \label{eq:Qn_def}
Q_k(t|\x_0) = \P_{\x_0}\{\N(t) = k\} = \lim\limits_{s\to 0} \frac{1}{k!} \partial_s^k G_s(t|\x_0).
\end{equation}
The knowledge of $G_s(t|\x_0)$ is thus equivalent to that of the
distribution of $\N(t)$.  By construction, the generating function
monotonously grows from $s = 0$ to $s=1$, at which $G_1(t|\x_0) = 1$
due to the normalization of probabilities $Q_k(t|\x_0)$.

\subsection{Integral equation}
\label{sec:integral}

We recall the probabilistic renewal-type argument from
\cite{Grebenkov26a} to establish an integral equation for the
generating function $G_s(t|\x_0)$.  For this purpose, we introduce two
random variables: the first-reaction time $\tau_{\rm a}$ on $\Gammaa$,
and the first-branching time $\tau_{\rm c}$ on $\Gammac$ (see Appendix
\ref{sec:Amodel} for their formal definitions).  Depending on which
of the three variables $t$, $\tau_{\rm c}$ or $\tau_{\rm a}$ is the
smallest, the generating function can be decomposed as
\begin{widetext}
\begin{align} \nonumber
G_s(t|\x_0) & = \E_{\x_0}\left\{ s^{\N(t)} 1_{t < \min\{\tau_{\rm c},\tau_{\rm a}\}} 
+ s^{\N(t)} 1_{\tau_{\rm a} < \min\{t,\tau_{\rm c}\}} 
+ s^{\N(t)} 1_{\tau_{\rm c} < \min\{t,\tau_{\rm a}\}}  \right\} \\  
\label{eq:Gs_decomposition}
& = s\, \P_{\x_0}\{t < \min\{\tau_{\rm c},\tau_{\rm a}\}\} + 
\P_{\x_0}\{ \tau_{\rm a} < \min\{t,\tau_{\rm c}\}\}
+ \E_{\x_0}\{ [G_s(t-\tau_{\rm c}|\X_{\tau_{\rm c}})]^2 1_{\tau_{\rm c} < \min\{t,\tau_{\rm a}\}}  \} .
\end{align}
\end{widetext}
In fact, if there is neither branching, nor absorption before $t$
(i.e., $t < \tau_{\rm c}$ and $t < \tau_{\rm a}$), the number of
particles at time $t$ remains to be equal to $1$, $\N(t) = 1$, that
gives the first term.  If the absorption event is the earliest (i.e.,
$\tau_{\rm a} < \min\{ t,\tau_{\rm c}\}$), the particle disappears,
$\N(t) = 0$, yielding the second term.  Finally, if the branching
event is the earliest, two newborn particles produce two independent
trees of descendants in the remaining time $t-\tau_{\rm c}$.  The
expectation over the descendants of each tree gives the factor
$G_s(t-\tau_{\rm c}|\X_{\tau_{\rm c}})$ in the third term (it is
squared because two trees are independent).  The remaining expectation
is over the first-branching time $\tau_{\rm c}$ and the position
$\X_{\tau_{\rm c}}$ of the branching event.

In order to perform this remaining average, one can employ the
single-particle propagator $P^{+}(\x,t|\x_0)$ satisfying the diffusion
equation with mixed boundary conditions:
\begin{subequations}  \label{eq:propagatorP}
\begin{align}  \label{eq:propagatorP_diff}
\partial_t P^{+}(\x,t|\x_0) - D \Delta P^{+}(\x,t|\x_0) & = 0 \quad \textrm{in}~\Omega, \\  \label{eq:PP_qc}
\partial_n P^{+}(\x,t|\x_0) + \qc P^{+}(\x,t|\x_0) & = 0 \quad \textrm{on}~ \Gammac, \\  \label{eq:PP_qa}
\partial_n P^{+}(\x,t|\x_0) + \qa P^{+}(\x,t|\x_0) & = 0 \quad \textrm{on}~ \Gammaa, \\
\partial_n P^{+}(\x,t|\x_0) & = 0 \quad \textrm{on}~ \Gammar, \\
P^{+}(\x,t=0|\x_0) &= \delta(\x-\x_0) .
\end{align}
\end{subequations}
Here $\Delta$ is the Laplace operator acting on $\x$, $\partial_n$ is
the normal derivative oriented outwards the domain $\Omega$, and
$\delta(\x-\x_0)$ is the Dirac distribution that fixes the starting
point at $\x_0$ at time $t=0$.  The superscript plus in
$P^{+}(\x,t|\x_0)$ highlights the {\it positive} rate $\qc$, which is
set on the catalytic region $\Gammac$; in other words, as we aim at
identifying which region would first affect the particle, both regions
$\Gammaa$ and $\Gammac$ are treated here as partially absorbing.  The
above propagator (or heat kernel) $P^{+}(\x,t|\x_0)$ is the most
common tool for characterizing competition between two partially
reactive traps in conventional diffusion-controlled reactions.  For a
single particle started from $\x_0$ at time $0$, $P^{+}(\x,t|\x_0)$ is
the probability density of finding that particle at a later time $t$
in a vicinity of the point $\x$, conditioned of not being destroyed on
$\Gammaa$ (with the rate $\qa$), nor on $\Gammac$ (with the rate
$\qc$).  In turn, the normal derivative of the propagator determines
the probability flux density on the boundary
\begin{equation}
j(\x,t|\x_0) = -D\partial_n P^{+}(\x,t|\x_0)  \qquad (\x\in\pa).
\end{equation}
When restricted to $\Gammac$, $j(\x,t|\x_0)$ is the joint probability
density of $\X_{\tau_{\rm c}}$ and $\tau_{\rm c}$, under the condition
that $\tau_{\rm c} < \tau_{\rm a}$.  In turn, the restriction on
$\Gammaa$ gives the joint probability density of $\X_{\tau_{\rm a}}$
and $\tau_{\rm a}$, under the condition that $\tau_{\rm a} < \tau_{\rm
c}$.  Moreover, the integral of $P^{+}(\x,t|\x_0)$ over all arrival
points is the survival probability up to time $t$:
\begin{equation}
S^+(t|\x_0) = \P_{\x_0}\{t < \min\{\tau_{\rm c},\tau_{\rm a}\}\} = \int\limits_{\Omega} d\x \, P^+(\x,t|\x_0).
\end{equation}
As a consequence, Eq. (\ref{eq:Gs_decomposition}) can be written as
\begin{align} \nonumber
& G_s(t|\x_0) = s S^+(t|\x_0) + \int\limits_0^t dt' \, \int\limits_{\Gammaa} d\x 
\, \qa D P^{+}(\x,t'|\x_0) \\  \label{eq:Gs_decomposition1}
& \qquad + \int\limits_0^t dt' \, \int\limits_{\Gammac} d\x \,\qc D P^{+}(\x,t'|\x_0) \, [G_s(t-t'|\x)]^2 ,
\end{align}
where we used the Robin boundary condition (\ref{eq:PP_qc},
\ref{eq:PP_qa}) on $\Gammaa$ and $\Gammac$ to express the joint
probability densities via the propagator.

It is convenient to rewrite the above equation in a slightly different
form.  For this purpose, let us first integrate
Eq. (\ref{eq:propagatorP_diff}) over $\x\in\Omega$ and use the
divergence theorem to get
\begin{align} \nonumber
\partial_{t'} \int\limits_{\Omega} d\x \, P^{+}(\x,t'|\x_0) & = - \qa D \int\limits_{\Gammaa} d\x \, P^{+}(\x,t'|\x_0) \\   \label{eq:identity_Pplus}
& - \qc D \int\limits_{\Gammac} d\x \, P^{+}(\x,t'|\x_0),
\end{align}
where we replaced $t$ by $t'$ and used again Robin boundary condition
on $\Gammaa$ and $\Gammac$.  Integrating this expression over $t'$
from $0$ to $t$ yields
\begin{align*}
& \qa D \int\limits_0^t dt' \int\limits_{\Gammaa} d\x \, P^{+}(\x,t'|\x_0) = 1 - 
\int\limits_{\Omega} d\x \, P^{+}(\x,t|\x_0) \\
& - \qc D \int\limits_0^t dt' \int\limits_{\Gammac} d\x \, P^{+}(\x,t'|\x_0).
\end{align*}
Substituting this relation into Eq. (\ref{eq:Gs_decomposition1}), we
obtain the following integral equation for the generating function:
\begin{align} \label{eq:Gs_integral}
& G_s(t|\x_0) = 1 - (1-s) S^{+}(t|\x_0)  \\  \nonumber
& \quad - \qc D \int\limits_0^t dt' \, \int\limits_{\Gammac} d\x \, P^{+}(\x,t'|\x_0) \, \bigl(1 - [G_s(t-t'|\x)]^2\bigr).
\end{align}
This equation, which is valid for any $t\geq 0$ and any $\x_0 \in
\overline{\Omega}$, is a particular case of more general integral
equation established in \cite{Grebenkov26a}.  If the propagator
$P^{+}(\x,t|\x_0)$ is known, one can restrict $\x_0$ to $\Gammac$,
solve this equation for $G_s(t|\x_0)\bigr|_{\Gammac}$, and then
reconstruct $G_s(t|\x_0)$ for any $\x_0 \in \Omega$ via
Eq. (\ref{eq:Gs_integral}).  As expected, the nonlinear form of the
right-hand side presents the major difficulty along this way.

The probabilistic renewal-type equation (\ref{eq:Gs_decomposition}),
which formalizes branching and absorption mechanisms in terms of the
associated first-reaction times $\tau_{\rm c}$ and $\tau_{\rm a}$, is
the key ingredient.  This description remains valid even for more
general Markov processes, including drifted and heterogeneous
diffusions \cite{Grebenkov26a}.  In fact, as the population size
changes exclusively on the boundary, the type of the stochastic motion
can only affect the distribution of $\tau_{\rm a}$ and the joint
distribution of $\tau_{\rm c}$ and $\X_{\tau_{\rm c}}$ through the
propagator $P^+(\x,t|\x_0)$.  Moreover, as the branching mechanism is
decoupled from the diffusive motion, other boundary-catalytic
processes can be incorporated.  For instance, if the particle splits
into $m$ offsprings, the second power in the last term of
Eq. (\ref{eq:Gs_decomposition}) has to be replaced by $m$.  One can
even consider heterogeneous branching when the number of offsprings
depends on the spatial location.  Despite the potential interest of
these extensions, we focus on ordinary diffusion and binary branching
in the following.

\subsection{PDE reformulation}
\label{sec:PDE}

Even though the integral equation (\ref{eq:Gs_integral}) fully
describes the generating function $G_s(t|\x_0)$, it is instructive to
transform this integral equation into an equivalent PDE problem
\cite{Grebenkov26a}:
\begin{subequations}  \label{eq:Gs_PDE}
\begin{align}  \label{eq:Gs_diff}
\partial_t G_s &= D \Delta G_s \quad \textrm{in}~\Omega, \\  \label{eq:Gs_qc}
\partial_n G_s &= \qc (G_s^2 - G_s) \quad \textrm{on}~ \Gammac, \\ \label{eq:Gs_qa}
\partial_n G_s &= \qa (1-G_s) \quad \textrm{on}~ \Gammaa, \\ \label{eq:Gs_qr}
\partial_n G_s &= 0 \quad \textrm{on}~ \Gammar, \\  \label{eq:Gs_ini}
G_s(0|\x_0) & = s,
\end{align}
\end{subequations}
where both operators $\Delta$ and $\partial_n$ act on the starting
point $\x_0$.  In fact, the integral equation (\ref{eq:Gs_integral})
can be recognized as a standard integral representation of the
solution of Eqs. (\ref{eq:Gs_PDE}), see Appendix \ref{sec:technical}.

Let us provide the intuitive interpretation of the initial-value
problem (\ref{eq:Gs_PDE}): (i) as absorption and branching events
occur exclusively on the boundary, ordinary diffusion in the bulk does
not affect the population size $\N(t)$, yielding the usual diffusion
equation (\ref{eq:Gs_diff}); (ii) on the catalytic region $\Gammac$,
one particle splits into two offsprings, so that the change in the
generating function is $G_s^2 - G_s$, with the catalytic rate $\qc$,
that gives the Robin-type boundary condition (\ref{eq:Gs_qc}); (iii)
similarly, the absorption event removes the particle and yields the
change $1 - G_s$, with the absorption rate $\qa$, in the Robin
boundary condition (\ref{eq:Gs_qa}) on $\Gammaa$; (iv) the remaining
reflecting region $\Gammar$ does not affect the population size, with
no change in the generating function and Neumann boundary condition
(\ref{eq:Gs_qr}); finally, (v) as the system starts from a single
particle, $\N(0) = 1$, one imposes the initial condition
(\ref{eq:Gs_ini}).  A more elaborate discussion of this probabilistic
interpretation in terms of boundary local times is given in Appendix
\ref{sec:proba}.

The {\it nonlinear} boundary condition (\ref{eq:Gs_qc}) is the key
difference with respect to conventional linear PDE descriptions (such
as Eqs. (\ref{eq:propagatorP})) used for characterizing
diffusion-controlled reactions and related first-reaction times.  This
nonlinearity is the reminiscent feature of branching processes.
However, most former works were dedicated to branching processes in
the bulk, so that nonlinear terms appeared in the diffusion equation,
e.g., in the Fisher-KPP equation
\cite{Fisher37,Kolmogorov37,Grindrod}.  In turn, the nonlinear term in
the boundary condition is the salient distinction of
boundary-catalytic branching processes.  Delmas and Vogt provided a
rigorous construction of such processes and highlighted their relation
to PDEs with nonlinear boundary conditions \cite{Delmas05} in the
steady-state regime \cite{Delmas05}.  In turn, our focus is the
time-dependent population dynamics and the asymptotic behavior of the
population size.

\subsection{Dual representations}
\label{sec:dual}

In the following, we aim at understanding the long-time behavior of
boundary-catalytic branching processes.  For this purpose, we deduce
another integral equation for the generating function $G_s(t|\x_0)$.
Setting
\begin{equation}  \label{eq:barGs_def}
\bar{G}_s(t|\x_0) = 1 - G_s(t|\x_0),
\end{equation}
we first rewrite the initial-value problem (\ref{eq:Gs_PDE}) as
\begin{subequations}  \label{eq:barGs}
\begin{align}
\partial_t \bar{G}_s &= D \Delta \bar{G}_s \quad \textrm{in}~\Omega, \\  \label{eq:barGs_qc}
\partial_n \bar{G}_s - \qc \bar{G}_s & =  - \qc \bar{G}_s^2 \quad \textrm{on}~ \Gammac, \\
\partial_n \bar{G}_s + \qa \bar{G}_s & = 0  \quad \textrm{on}~ \Gammaa, \\
\partial_n \bar{G}_s &= 0 \quad \textrm{on}~ \Gammar, \\
\bar{G}_s(0|\x_0) & = 1-s. 
\end{align}
\end{subequations}
This elementary modification removed the inhomogeneous term from the
Robin boundary condition on $\Gammaa$.  The curious point of this
representation is that the catalytic rate $\qc$ in the left-hand side
of Eq. (\ref{eq:barGs_qc}) appears with the {\it negative} sign, in
sharp contrast with the conventional positive sign of the absorption
rate $\qa$.

To exploit this feature, we introduce another single-particle
propagator, denoted as $P^-(\x,t|\x_0)$, which satisfies
\begin{subequations}  \label{eq:propagator}
\begin{align}  \label{eq:propagator_diff}
\partial_t P^{-}(\x,t|\x_0) - D \Delta P^{-}(\x,t|\x_0) & = 0 \quad \textrm{in}~\Omega, \\  \label{eq:P_qc}
\partial_n P^{-}(\x,t|\x_0) - \qc P^{-}(\x,t|\x_0) & = 0 \quad \textrm{on}~ \Gammac, \\  \label{eq:P_qa}
\partial_n P^{-}(\x,t|\x_0) + \qa P^{-}(\x,t|\x_0) & = 0 \quad \textrm{on}~ \Gammaa, \\
\partial_n P^{-}(\x,t|\x_0) & = 0 \quad \textrm{on}~ \Gammar, \\
P^{-}(\x,t=0|\x_0) &= \delta(\x-\x_0) .
\end{align}
\end{subequations}
The only but crucial difference from the former propagator
$P^+(\x,t|\x_0)$ is the negative sign in front of the catalytic rate
$\qc$ (as highlighted by the superscript minus).  In Appendix
\ref{sec:Pminus}, we argue that $P^-(\x,t|\x_0)$ is the mean
population density in $\x$ at time $t$. 
Considering the right-hand side of Eq. (\ref{eq:barGs_qc}) as a ``flux
source'' on $\Gammac$, one can express the solution of the
initial-value problem (\ref{eq:barGs}) as (see Appendix
\ref{sec:technical} for details):
\begin{align} \label{eq:barGs_int}
\bar{G}_s(t|\x_0) & = (1-s) S^-(t|\x_0) - \qc D \int\limits_0^t dt' \\ \nonumber
& \times \int\limits_{\Gammac} d\x \, P^-(\x,t-t'|\x_0)\, \bar{G}_s^2(t'|\x),
\end{align}
where
\begin{equation}
S^-(t|\x_0) = \int\limits_{\Omega} d\x \, P^-(\x,t|\x_0).
\end{equation}
In this way, we deduced a dual integral representation for the
generating function $G_s(t|\x_0)$ in terms of the propagator
$P^-(\x,t|\x_0)$.  The dual nature can be more explicitly seen by
rewriting the former integral equation (\ref{eq:Gs_integral}) for
$\bar{G}_s(t|\x_0)$ as
\begin{align} \label{eq:barGs_int2}  
\bar{G}_s(t|\x_0) & = (1-s) S^{+}(t|\x_0) - \qc D \int\limits_0^t dt'  \\  \nonumber 
& \times \int\limits_{\Gammac} d\x \, P^{+}(\x,t-t'|\x_0) 
\biggl(\bar{G}_s^2(t'|\x) - 2\bar{G}_s(t'|\x)\biggr).
\end{align}

Comparing these two expressions, one can appreciate the flexibility of
handling the Robin boundary condition via the integral term.  In fact,
even though the integral over $\Gammac$ was originally needed to
incorporate the quadratic term in the boundary condition
(\ref{eq:barGs_qc}), it can also be used to accommodate the linear
term $\qc \bar{G}_s$ as well.  In other words, one has a freedom in
choosing a suitable propagator by moving the linear term from the
boundary condition to the integral term.

An immediate consequence of this freedom is yet another
representation:
\begin{align} \label{eq:barGs_P0}
\bar{G}_s(t|\x_0) & = (1-s) S(t|\x_0) - \qc D \int\limits_0^t dt' \\    \nonumber 
& \times \int\limits_{\Gammac} d\x \, P(\x,t-t'|\x_0) 
\biggl(\bar{G}_s^2(t'|\x) - \bar{G}_s(t'|\x)\biggr),
\end{align}
with the third propagator $P(\x,t|\x_0)$ satisfying
\begin{subequations}  \label{eq:propagator0}
\begin{align}  \label{eq:propagator0_diff}
\partial_t P(\x,t|\x_0) - D \Delta P(\x,t|\x_0) & = 0 \quad \textrm{in}~\Omega, \\  \label{eq:P0_qc}
\partial_n P(\x,t|\x_0) & = 0 \quad \textrm{on}~ \Gammac, \\  \label{eq:P0_qa}
\partial_n P(\x,t|\x_0) + \qa P(\x,t|\x_0) & = 0 \quad \textrm{on}~ \Gammaa, \\
\partial_n P(\x,t|\x_0) & = 0 \quad \textrm{on}~ \Gammar, \\
P(\x,t=0|\x_0) &= \delta(\x-\x_0) ,
\end{align}
\end{subequations}
and
\begin{equation}  \label{eq:S0}
S(t|\x_0) = \int\limits_{\Omega} d\x \, P(\x,t|\x_0).
\end{equation}
In this representation, the propagator itself corresponds to a
setting, in which the catalytic region $\Gammac$ is treated as
reflecting, whereas the catalytic effect is incorporated into
$\bar{G}_s(t|\x_0)$ through the integral over $\Gammac$ in
Eq. (\ref{eq:barGs_P0}).  The major advantage of this presentation is
that $P(\x,t|\x_0)$ does not depend on $\qc$.

In the following, we will exploit the advantages of all three
representations.  In particular, Eqs. (\ref{eq:barGs_int}) and
(\ref{eq:barGs_int2}) will be used in Sec. \ref{sec:Gs_asympt} for the
long-time asymptotic analysis, whereas the integral equation
(\ref{eq:barGs_P0}) turns out to be more suitable for a numerical
solution (see Appendices \ref{sec:annulus} and \ref{sec:numerics}).

\subsection{Distribution of the population size}
\label{sec:distrib}

According to Eq. (\ref{eq:Qn_def}), the generating function
$G_s(t|\x_0)$ allows one to determine the probability $Q_k(t|\x_0)$ of
having $k$ particles at time $t$ via $k$-fold differentiation with
respect to $s$.  However, a numerical computation of these derivatives
may be challenging, especially for large $k$.  It is therefore
convenient to provide an alternative way to access the probabilities
$Q_k(t|\x_0)$.  The probability $Q_0(t|\x_0)$ is obtained by setting
$s = 0$ to $G_s(t|\x_0)$, i.e., one can use any of the above
representations for $G_s(t|\x_0)$.  For instance,
Eq. (\ref{eq:Gs_integral}) reads
\begin{align} \label{eq:Q0_integral}
& Q_0(t|\x_0) = 1 - S^{+}(t|\x_0)  \\  \nonumber
& \quad + \qc D \int\limits_0^t dt' \int\limits_{\Gammac} d\x \, P^{+}(\x,t'|\x_0) \, \bigl([Q_0(t-t'|\x)]^2 - 1\bigr) .
\end{align}

For $k > 0$, one can differentiate the integral equation
(\ref{eq:Gs_integral}) $k$ times with respect to $s$ and evaluate it
at $s = 0$ to get
\begin{align} \label{eq:Qn_integral}
Q_k(t|\x_0) & = S^+(t|\x_0) \delta_{k,1} \\  \nonumber
& \quad + \qc D \int\limits_0^t dt'  \int\limits_{\Gammac} d\x \, P^{+}(\x,t'|\x_0) \, H_k(t-t'|\x) ,
\end{align}
where
\begin{equation}  \label{eq:Hn_def}
H_k(t|\x) = \frac{1}{k!} \lim\limits_{s\to 0} \partial^k_s [G_s(t|\x)]^2 = \sum\limits_{j=0}^k Q_j(t|\x) Q_{k-j}(t|\x)
\end{equation}
is obtained by using the general Leibniz rule to differentiate the
product of two functions (note that the binomial coefficient is
compensated by the factorials that appear from the definition
(\ref{eq:Qn_def}) of $Q_j(t|\x_0)$).  Note that the function
$H_k(t|\x)$ is linear with respect to $Q_k(t|\x)$, i.e., it does not
include quadratic terms like $Q_k^2(t|\x)$.  As a consequence,
Eq. (\ref{eq:Qn_integral}) is actually linear in $Q_k(t|\x_0)$, in
sharp contrast to the nonlinear integral equation
(\ref{eq:Q0_integral}) for $Q_0(t|\x_0)$.  Once $Q_0(t|\x_0)$ is
found, the probabilities $Q_1(t|\x_0), Q_2(t|\x_0),
\ldots$ can be obtained by solving iteratively the linear integral
equations (\ref{eq:Qn_integral}) for $k = 1,2,\ldots$. 

As for the generating function, one can get alternative integral
equations for $Q_k(t|\x_0)$ based on the propagators $P^-(\x,t|\x_0)$
or $P(\x,t|\x_0)$.  For instance, the $k$-fold derivative of the
integral equation (\ref{eq:barGs_P0}) with respect to $s$, evaluated
at $s = 0$, yields for $k > 0$
\begin{align} \nonumber
Q_k(t|\x_0) & = S(t|\x_0) \delta_{k,1} + \qc D \int\limits_0^t dt' \int\limits_{\Gammac} d\x \, P(\x,t'|\x_0) \\ \label{eq:Qn_P0}
& \times \biggl(H_k(t-t'|\x) - Q_k(t-t'|\x)\biggr),
\end{align}
and
\begin{align} \nonumber
\bar{Q}_0(t|\x_0) & = S(t|\x_0) + \qc D \int\limits_0^t dt' \int\limits_{\Gammac} d\x \, P(\x,t'|\x_0) \\ \label{eq:barQ0_P0}
& \times \biggl(\bar{Q}_0(t-t'|\x) - \bar{Q}_0^2(t-t'|\x)\biggr),
\end{align}
where $\bar{Q}_0(t|\x_0) = 1 - Q_0(t|\x_0)$.  Setting $\x_0 \in
\Gammac$, the latter integral equation determines $\bar{Q}_0(t|\x_0)$.

Yet another description is obtained by applying the $k$-fold
derivative to the PDE problem (\ref{eq:Gs_PDE}) to get for any $k >
0$:
\begin{subequations}  \label{eq:Qn_PDE}
\begin{align}  
\partial_t Q_k - D \Delta Q_k & = 0 \quad \textrm{in}~\Omega, \\  \label{eq:Qn_qc}
\partial_n Q_k + \qc Q_k &= \qc H_k  \quad \textrm{on}~ \Gammac, \\ 
\partial_n Q_k + \qa Q_k &= 0 \quad \textrm{on}~ \Gammaa, \\ 
\partial_n Q_k &= 0 \quad \textrm{on}~ \Gammar, \\  
Q_k(0|\x_0) & = \delta_{k,1}.
\end{align}
\end{subequations}
Once again, this PDE problem is linear in $Q_k$ and can be solved
iteratively for $Q_1$, $Q_2$, etc., once $Q_0$ is found.

\section{Long-time behavior of the generating function}
\label{sec:Gs_asympt}

In this section, we discuss the long-time asymptotic behavior of the
generating function.  Since $G_1(t|\x_0) \equiv 1$ due to the
probability normalization, we exclude $s = 1$ below by considering $0
\leq s < 1$.  For convenience, we will inspect the complementary
function $\bar{G}_s(t|\x_0)$ given by Eq. (\ref{eq:barGs_def}).  We
first introduce three distinct regimes and then derive the asymptotic
behavior of $\bar{G}_s(t|\x_0)$ separately in each regime.

\subsection{Three asymptotic regimes}
\label{sec:regimes}

The negative sign of the catalytic rate in the boundary condition
(\ref{eq:P_qc}) drastically changes the asymptotic behavior of the
propagator $P^-(\x,t|\x_0)$ as compared to $P^+(\x,t|\x_0)$.  In fact,
when both absorbing and catalytic regions were treated as killing
(with the positive rates $\qc$ and $\qa$), there was no branching, and
the particle had to disappear.  This resulted in the exponential decay
of the propagator $P^+(\x,t|\x_0)$ as $t\to \infty$, as in
conventional diffusion-controlled reactions.  In turn, the propagator
$P^-(\x,t|\x_0)$ describes the diffusive dynamics, in which the
catalytic region $\Gammac$ produces new particles via branching
events, i.e., $\Gammac$ plays the role of a source \cite{Grebenkov26}.
If the production of particles via branching events is stronger than
their absorption on $\Gammaa$, the population size grows.

Since the domain $\Omega$ is bounded with a smooth boundary, the
Laplace operator governing the diffusive dynamics has a discrete
spectrum, despite the presence of the catalytic region with negative
reactivity \cite{Levitin}.  In other words, there are infinitely many
eigenpairs $\{\lambda_k^-, u_k^-(\x)\}$ satisfying
\begin{subequations}  \label{eq:Laplace}
\begin{align}
-\Delta u_k^- & = \lambda_k^- u_k^-  \quad \textrm{in}~\Omega, \\
\partial_n u_k^- - \qc u_k^- & = 0 \quad \textrm{on}~ \Gammac, \\ 
\partial_n u_k^- + \qa u_k^- & = 0 \quad \textrm{on}~ \Gammaa, \\
\partial_n u_k^- & = 0 \quad \textrm{on}~ \Gammar. 
\end{align}
\end{subequations}
The eigenvalues are enumerated by index $k$ in the increasing order,
\begin{equation}
\lambda_0^- < \lambda_1^- \leq \lambda_2^- \leq \ldots \nearrow +\infty ,
\end{equation} 
whereas the associated eigenfunctions $\{u_k^-\}$ form a complete
orthonormal basis of the space $L^2(\Omega)$ of square-integrable
functions on $\Omega$, i.e.,
\begin{equation}
\int\limits_{\Omega} d\x \, u_j^-(\x) \, u_k^-(\x) = \delta_{j,k} .
\end{equation}
Since the domain $\Omega$ is connected (as we always assume), the
principal (smallest) eigenvalue $\lambda_0^-$ is simple (of
multiplicity $1$), whereas the associated eigenfunction $u_0^-$ does
not change the sign in $\Omega$ and can thus be chosen to be positive
(see, e.g., \cite{Hassannezhad24}).  
As the propagator admits the spectral expansion over this eigenbasis,
\begin{equation}  \label{eq:p_spectral}
P^-(\x,t|\x_0) = \sum\limits_{k=0}^\infty u_k^-(\x) \, u_k^-(\x_0) \, e^{-Dt\lambda_k^-} ,
\end{equation}
the principal eigenvalue $\lambda_0^-$ determines the long-time
asymptotic behavior of the propagator and the related quantities (such
as $S^-(t|\x_0)$).

\begin{figure}
\begin{center}
\includegraphics[width=85mm]{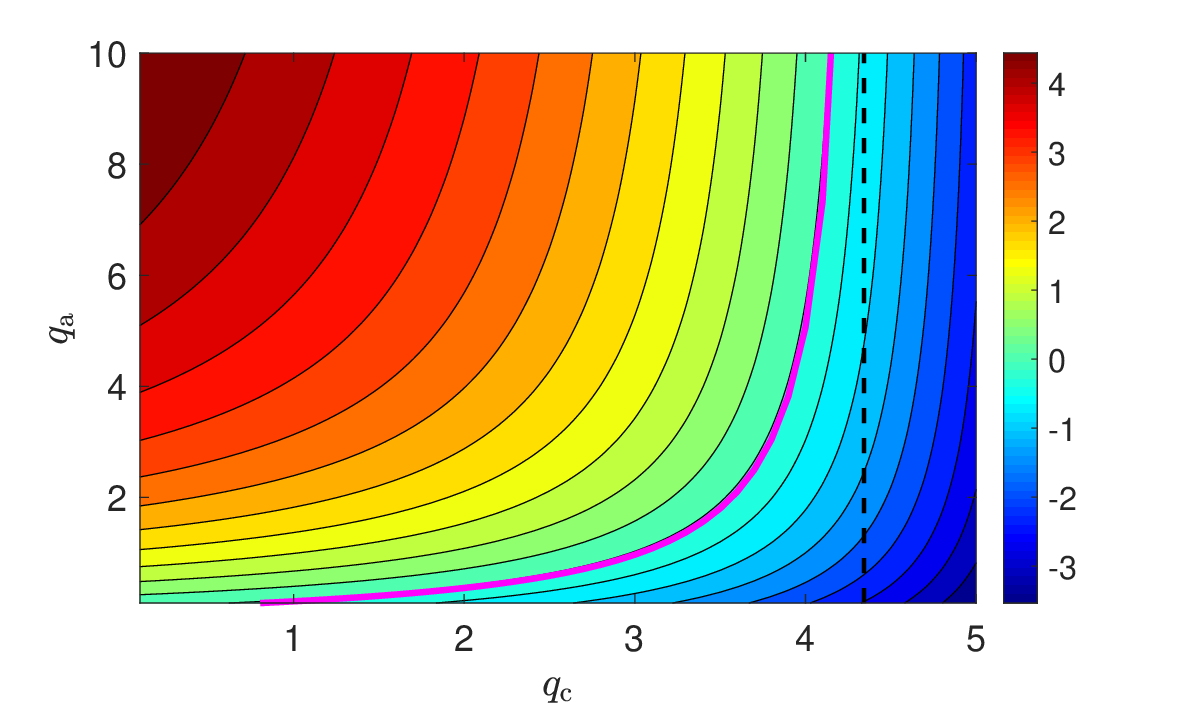} % {lam0_surf.eps}
\end{center}
\caption{
Contour plot of the principal eigenvalue $\lambda_0^-$ as a function
of $\qc$ and $\qa$ for the circular annulus with $R = 0.1$ and $L =
1$.  The vertical dashed line presents $\qccrit$ from
Eq. (\ref{eq:qccrit}), while thick magenta line shows the critical
line $\qc = \mu_0(\qa)$ that separates the subcritical regime
($\lambda_0^- > 0$, top left) and the supercritical regime
($\lambda_0^- < 0$, bottom right).  The function $\mu_0(\qa)$ is given
by Eq. (\ref{eq:mu0}). }
\label{fig:lam0}
% [lam0, qc,qa] = A_Yilin4_annulus_lam1_surf();
\end{figure}

According to the variational principle, one can represent the
principal eigenvalue as
\begin{equation}  \label{eq:lambda0_var}
\lambda_0^- = \inf\limits_{u\not\equiv 0} \biggl\{\frac{\int\nolimits_{\Omega} d\x \, |\nabla u|^2 
+ \qa \int\nolimits_{\Gammaa} d\x \, u^2 - \qc \int\nolimits_{\Gammac} d\x \, u^2}{\int\nolimits_{\Omega} d\x \, u^2}
\biggr\},
\end{equation}
where the infimum is taken over all functions $u\not\equiv 0$ from a
suitable functional space $H^1(\Omega)$ \cite{Levitin}.  Let us
comment on several particular situations.  (i) When there is no
catalytic region (i.e., $\Gammac = \emptyset$ and $\qa > 0$), the last
term in the numerator of Eq. (\ref{eq:lambda0_var}) disappears, while
the remaining terms are positive.  As a consequence, the principal
eigenvalue is strictly positive, $\lambda_0^- > 0$ \cite{Levitin},
ensuring the exponential decay of the propagator at long times, as
expected for conventional diffusion-controlled reactions.  (ii) In the
special case of no reactivity ($\Gammac = \Gammaa = \emptyset$), the
infimum is achieved on a constant function, so that $\lambda_0^- = 0$,
and the system reaches a steady-state limit.  Moreover, setting the
test function $u$ to be a constant, one gets an elementary upper bound
in the general situation:
\begin{equation}  \label{eq:lambda0_upper}
\lambda_0^- \leq \frac{\qa |\Gammaa| - \qc |\Gammac|}{|\Omega|} \,.
\end{equation}
(iii) In the purely catalytic regime (i.e., $\qc > 0$ and $\Gammaa =
\emptyset$), this inequality implies that the principal eigenvalue is
strictly negative, yielding an exponential growth of the population.
In general, however, the absorption and branching events compete with
each other, and the sign of the principal eigenvalue $\lambda_0^-$
depends on both rates $\qa$ and $\qc$, as well as on the shape of the
domain and on the geometric properties of the absorbing and catalytic
regions.  
Figure \ref{fig:lam0} illustrates the behavior of the principal
eigenvalue $\lambda_0^-$ as a function of $\qc$ and $\qa$ for the
circular annulus.  

The possibility of compensating branching events on $\Gammac$ by
absorption events on $\Gammaa$ for the mean population size
$N_1(t|\x_0) = \E_{\x_0}\{ \N(t)\}$ was investigated in
\cite{Grebenkov26}; in particular, it was shown when and how the
absorption rate $\qa$ can be chosen to ensure the condition
$\lambda_0^- = 0$.  Three long-time asymptotic regimes were
distinguished: (i) the subcritical extinction regime ($\lambda_0^- >
0$) when $N_1(t|\x_0)$ vanishes exponentially, (ii) the critical
regime ($\lambda_0^- = 0$) when $N_1(t|\x_0)$ reaches a steady-state
limit, and (iii) the supercritical growth regime ($\lambda_0^- < 0$),
in which $N_1(t|\x_0)$ grows exponentially fast.  In the following, we
provide a quantitative description of the long-time asymptotic
behavior of the generating function in the three regimes.

\subsection{Subcritical regime}
\label{eq:barGs_ext}

In the subcritical regime ($\lambda_0^- > 0$), both $P^-(\x,t|\x_0)$
and $S^-(t|\x_0)$ decay exponentially as $e^{-Dt\lambda_0^-}$ so that
$\bar{G}_s(t|\x_0)$, obeying Eq. (\ref{eq:barGs_int}), is expected to
inherit this decay.  To analyze this behavior, we use the Laplace
transform, which is defined for a function $f(t)$ as
\begin{equation*}
\L\{ f(t)\}(p) = \tilde{f}(p) = \int\limits_0^\infty dt \, e^{-pt} \, f(t) ,
\end{equation*}
where both tilde and $\L$ notations are introduced.  The application
of the Laplace transform removes convolution in time in the second
term of Eq. (\ref{eq:barGs_int}):
\begin{align}  \label{eq:barGs_Laplace} 
\L\{ \bar{G}_s(t|\x_0)\} (p) & = (1-s) \tilde{S}^-(p|\x_0)  \\  \nonumber
& - \qc D \int\limits_{\Gammac} d\x \, \tilde{P}^-(\x,p|\x_0) \, \L\{ \bar{G}_s^2(t|\x)\}(p).
\end{align}
According to the spectral expansion (\ref{eq:p_spectral}), the Laplace
transforms $\tilde{S}^-(p|\x_0)$ and $\tilde{P}^-(\x,p|\x_0)$ have the
poles at $p_n = - D\lambda_n^-$, and the simple largest pole $p_0$
controls the long-time asymptotic behavior.  If $\bar{G}_s(t|\x_0)$
decays as $e^{-Dt\lambda_0^-}$ as $t\to \infty$, the largest pole of
$\L\{ \bar{G}_s^2(t|\x)\}(p)$ is $-2D\lambda_0^-$, so that this
function has no singularity at $p_0$.  As a consequence, the
right-hand side of Eq. (\ref{eq:barGs_Laplace}) has a simple pole at
$p = p_0$ and can thus be written near $p_0$ as
\begin{align}  \label{eq:barGs_Laplace2} 
& \L\{ \bar{G}_s(t|\x_0)\} (p) \simeq \frac{1}{p-p_0} \biggl[(1-s) C_0^- u_0^-(\x_0)  \\  \nonumber
& - \qc D u_0^-(\x_0) \int\limits_{\Gammac} d\x \, u_0^-(\x) \, \L\{ \bar{G}_s^2(t|\x)\}(p_0)\biggr],
\end{align}
where
\begin{equation}  \label{eq:Ck_def}
C_k^- = \int\limits_{\Omega} d\x \, u_k^-(\x).
\end{equation}
The inverse Laplace transform of this relation yields the long-time
asymptotic behavior
\begin{equation}  \label{eq:barGs_asympt_ext}
\bar{G}_s(t|\x_0) \simeq g_s \, C_0^- \, u_0^-(\x_0)\, e^{-Dt\lambda_0^-}  \qquad (t\to \infty),
\end{equation}
with the dimensionless coefficient $g_s$ given by
\begin{equation}  \label{eq:barGs_asympt_a}
g_s = 1-s  - \frac{\qc D}{C_0^-} \int\limits_{\Gammac} d\x \, u_0^-(\x) \int\limits_0^\infty dt \, e^{Dt\lambda_0^-} \bar{G}_s^2(t|\x).
\end{equation}
This representation is rather formal because it requires the knowledge
of $\bar{G}_s(t|\x)$ for all times $t$.  At the same time, the
asymptotic relation (\ref{eq:barGs_asympt_ext}) captures correctly the
dependence of $\bar{G}_s(t|\x_0)$ on both $\x_0$ and $t$ in the
long-time regime.

To illustrate the behavior of $\bar{G}_s(t|\x_0)$, we consider again
diffusion in the circular annulus $\Omega = \{\x\in\R^2 ~:~ R < |\x| <
L\}$ and restrict our attention to the case $s = 0$, for which case
$\bar{G}_0(t|\x_0) = \bar{Q}_0(t|\x_0)$.  Moreover, we fix the
starting point $\x_0$ to be on the catalytic region $\Gammac$ so that
$|\x_0| = R$, and use the shortcut notation $\bar{Q}_0(t|\x_0) =
\bar{Q}_0(t|R)$ for the probability that at least one particle is
present at time $t$ when started from the catalytic region $\Gammac$
(similar results for other values of $s$ and $\x_0$ are not shown).
We compute this function by solving numerically the integral equation
(\ref{eq:barGs_P0}), as described in Appendix
\ref{sec:numerics}.  

Figure \ref{fig:Q0_asympt2}(a) presents the behavior of
$\bar{Q}_0(t|R)$ in the subcritical regime with $\qc \approx 3.91$ (we
used the same parameters as in Fig. \ref{fig:Nt_traj}(a)).  As
described above, the probability $\bar{Q}_0(t|R)$ decreases
exponentially fast, with the rate $D\lambda_0^-$, determined by the
principal eigenvalue $\lambda_0^-$ (see Appendix
\ref{sec:annulus_eigen} for its numerical computation).  By evaluating
numerically the integral in Eq. (\ref{eq:barGs_asympt_a}) with the
already constructed numerical solution $\bar{G}_0(t|R)$, we managed to
get the correct amplitude in the asymptotic relation
(\ref{eq:barGs_asympt_ext}), as shown by dashed line.  Monte Carlo
simulations (crosses) described in Appendix \ref{sec:MC} are in
perfect agreement with the numerical solution.

\begin{figure}[!ht]
\begin{center}
\includegraphics[width=0.99\columnwidth]{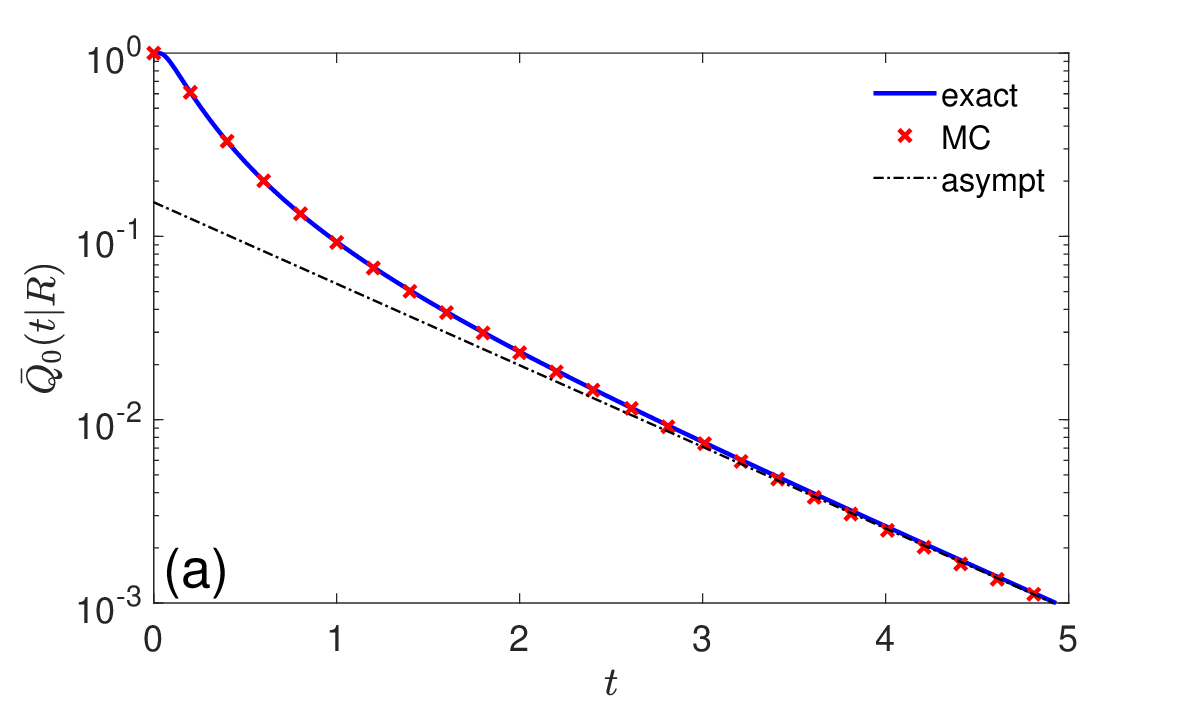} % {Q0R_qc09_new.eps}
\includegraphics[width=0.99\columnwidth]{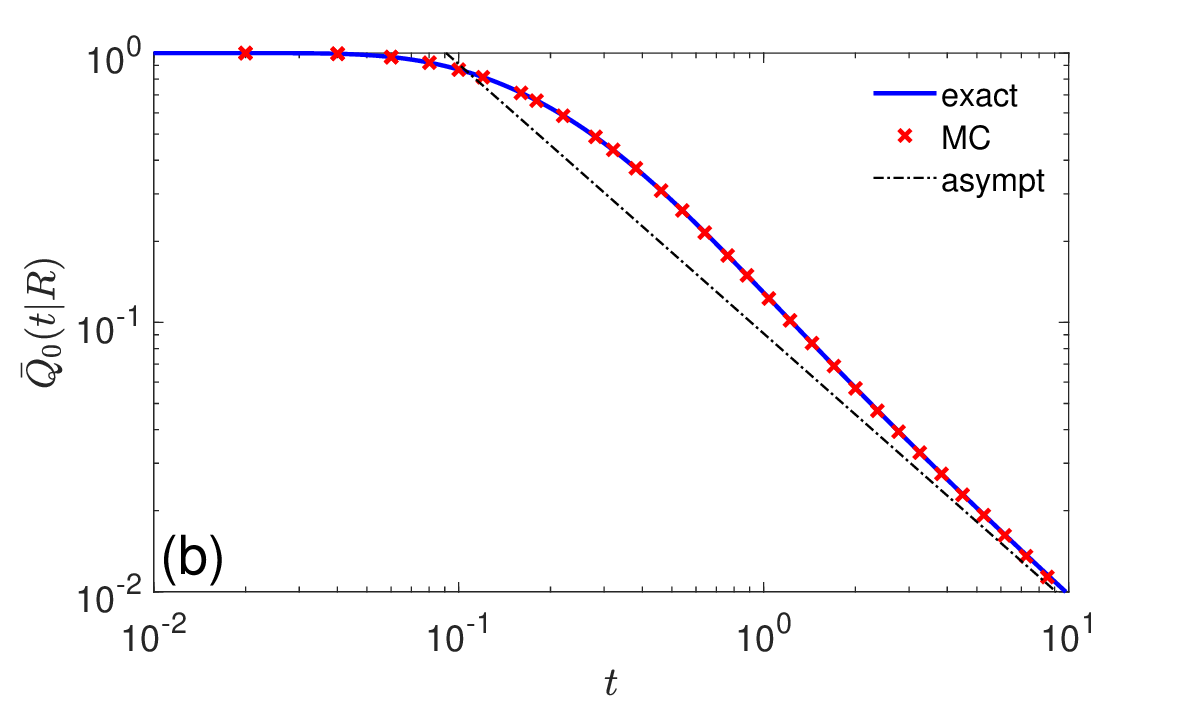} % {Q0R_qc10_new.eps}
\includegraphics[width=0.99\columnwidth]{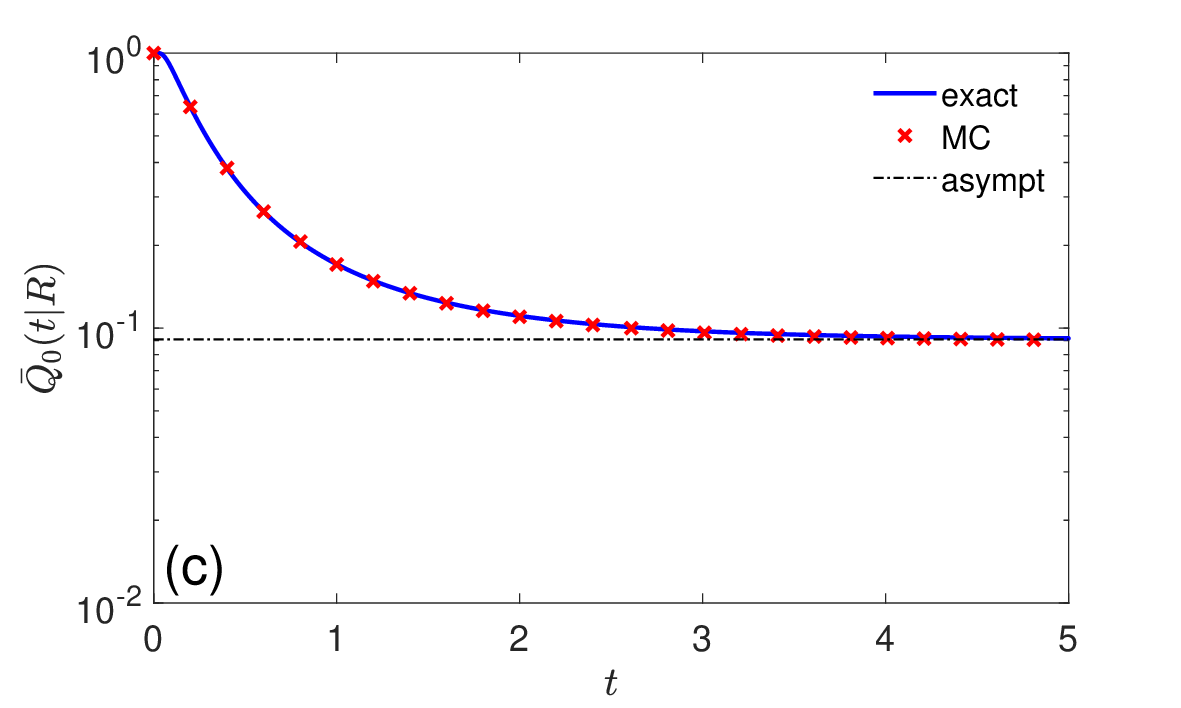} % {Q0R_qc11_new.eps}
\end{center}
\caption{
The probability $\bar{Q}_0(t|R) = \bar{G}_0(t|R)$ as a function of
time $t$ in the circular annulus with $R = 0.1$, $L = 1$, $D = 1$,
$\qa = \infty$.  Three panels illustrate three regimes: {\bf (a)}
subcritical ($q_{\c} = 0.9\, \qccrit \approx 3.91$); {\bf (b)}
critical ($q_{\c} = \qccrit \approx 4.34$); and {\bf (c)}
supercritical ($q_{\c} = 1.1\, \qccrit
\approx 4.78$), with $\qccrit = 1/(R\ln(L/R))$.  The starting point
$\x_0$ is set on the catalytic region (inner circle).  Solid line
presents a numerical solution of the integral equation
(\ref{eq:barGs_P0}) as described in Appendix \ref{sec:numerics};
crosses show an estimation from Monte Carlo simulations with $10^6$
particles (see Appendix \ref{sec:MC}); dashed line indicates the
asymptotic relations (\ref{eq:barGs_asympt_ext},
\ref{eq:barGs_asympt_crit}, \ref{eq:Gs_asympt_growth}) in
subcritical, critical, and supercritical regimes, respectively, with
the limit $\bar{G}_0(\infty|R)$ given by
Eq. (\ref{eq:barGsR_inf_circular}) in the supercritical regime.}
\label{fig:Q0_asympt2}
% [Q0,t] = A_Yilin4_simu7new2(2);
% [Q0,t] = A_Yilin4_simu7new2(1);
% [Q0,t] = A_Yilin4_simu7new2(3);   % I still use 1e5 particles in this case
\end{figure}

\subsection{Critical regime}

In the critical regime, $S^-(t|\x_0)$ and $P^-(\x,t|\x_0)$ approach
constants.  Taking the time derivative of Eq. (\ref{eq:barGs_int}), we
get then at long times
\begin{equation*}
\partial_t \bar{G}_s(t|\x_0) \simeq - \qc D  \int\limits_{\Gammac} d\x \, u_0^-(\x) \, u_0^-(\x_0)\, \bar{G}_s^2(t|\x).
\end{equation*}
One sees that the space dependence is given again by $u_0^-(\x_0)$,
whereas the remaining ordinary differential equation with respect to
time can be solved exactly, yielding
\begin{equation}  \label{eq:barGs_asympt_crit}
\bar{G}_s(t|\x_0) \simeq \frac{C_0^- u_0^-(\x_0)}{q_0 \qc Dt}   \qquad (t\to \infty),
\end{equation}
with the characteristic inverse length
\begin{equation}   \label{eq:q0}
q_0 = C_0^- \int\limits_{\Gammac} d\x\, [u_0^-(\x)]^3.
\end{equation}
This asymptotic behavior is illustrated on
Fig. \ref{fig:Q0_asympt2}(b).
As the next-order term in Eq. (\ref{eq:barGs_asympt_crit}) may give
the timescale for the applicability of this asymptotic relation, its
computation presents an interesting perspective.

\subsection{Supercritical regime}
\label{sec:Gs_long_growth}

In the supercritical regime, both $P^-(\x,t|\x_0)$ and $S^-(t|\x_0)$
grow exponentially as $t\to \infty$.  In analogy to the subcritical
regime, one might naively expect that the structure of the integral
equation (\ref{eq:barGs_int}) would imply an exponential growth of
$\bar{G}_s(t|\x_0)$.  However, this function must remain bounded
between $0$ and $1$, so that such a growth is not possible.  Even
though this equation remains valid, it is not suitable for the
analysis of the supercritical regime, and we switch to
Eq. (\ref{eq:barGs_int2}), which involves the decaying functions
$P^+(\x,t|\x_0)$ and $S^+(t|\x_0)$.  This propagator also admits the
spectral expansion
\begin{equation}  \label{eq:pp_spectral}
P^+(\x,t|\x_0) = \sum\limits_{k=0}^\infty u_k^+(\x) \, u_k^+(\x_0) \, e^{-Dt\lambda_k^+} ,
\end{equation}
where $\lambda_k^+$ and $u_k^+$ are the eigenvalues and eigenfunctions
of the Laplace operator with the positive rate $\qc$ on $\Gammac$:
\begin{subequations}  \label{eq:Laplace0}
\begin{align}
-\Delta u_k^+ & = \lambda_k^+ u_k^+  \quad \textrm{in}~\Omega, \\
\partial_n u_k^+ + \qc u_k^+ & = 0 \quad \textrm{on}~ \Gammac, \\ 
\partial_n u_k^+ + \qa u_k^+ & = 0 \quad \textrm{on}~ \Gammaa, \\
\partial_n u_k^+ & = 0 \quad \textrm{on}~ \Gammar. 
\end{align}
\end{subequations}
As previously, the eigenvalues are enumerated by index $k$ in the
increasing order,
\begin{equation}
0< \lambda_0^+ < \lambda_1^+ \leq \lambda_2^+ \leq \ldots \nearrow +\infty ,
\end{equation} 
whereas the associated eigenfunctions $\{u_k^+\}$ form a complete
orthonormal basis of the space $L^2(\Omega)$:
\begin{equation}
\int\limits_{\Omega} d\x \, u_j^+(\x) \, u_k^+(\x) = \delta_{j,k} .
\end{equation}
When the domain $\Omega$ is connected (as we always assume), the
principal (smallest) eigenvalue $\lambda_0^+$ is simple (of
multiplicity $1$) and strictly positive due to the positive
reactivities on both $\Gammaa$ and $\Gammac$.  This expansion
immediately implies that the principal eigenvalue $\lambda_0^+$
determines the long-time asymptotic decay of the associated propagator
and the related quantities.

Since $S^+(t|\x_0)$ vanishes in the long-time limit, it is evident
from Eq. (\ref{eq:barGs_int2}) that $\bar{G}_s \equiv 0$ is always a
solution in the limit $t\to \infty$.  Is it possible to get a nonzero
steady-state solution?  Let us first assume that such a nontrivial
solution exists and then inspect the conditions for its existence.
Replacing $\bar{G}_s(t|\x_0)$ by its limit $\bar{G}_s(\infty|\x_0)$ in
Eq. (\ref{eq:barGs_int2}), we get in the leading order
\begin{equation}  \label{eq:Gs_infty}
\bar{G}_s(\infty|\x_0) = \qc \int\limits_{\Gammac} d\x \, \G^+(\x,\x_0) \bigl(2\bar{G}_s(\infty|\x) - \bar{G}_s^2(\infty|\x)\bigr),
\end{equation}
where
\begin{equation}  \label{eq:GLaplace_spectral}
\G^+(\x,\x_0) = \int\limits_0^\infty dt' \, D \, P^+(\x,t'|\x_0) = \sum\limits_{k=0}^\infty  \frac{u_k^+(\x_0) u_k^+(\x)}{\lambda_k^+}
\end{equation}
is the Green's function of the Laplace operator satisfying
\begin{subequations}  \label{eq:GLaplace_PDE}
\begin{align}
-\Delta \G^+(\x,\x_0) & = \delta(\x-\x_0) \quad \textrm{in}~\Omega, \\
\partial_n \G^+(\x,\x_0) + \qc \G^+(\x,\x_0) & = 0 \quad \textrm{on}~ \Gammac, \\ 
\partial_n \G^+(\x,\x_0) + \qa \G^+(\x,\x_0) & = 0 \quad \textrm{on}~ \Gammaa, \\
\partial_n \G^+(\x,\x_0) & = 0 \quad \textrm{on}~ \Gammar. 
\end{align}
\end{subequations}

Since $\G^+(\x,\x_0)$ is independent $s$, the solution of the integral
equation (\ref{eq:Gs_infty}) does not also depend on $s$.  Given that
$\bar{G}_1(t|\x_0) \equiv 0$ due to the probability distribution
normalization, one might expect that $\bar{G}_s(\infty|\x_0) \equiv 0$
for any $s$.  We will argue below that this conclusion is valid in the
subcritical and critical regimes.  In turn, a strictly positive
solution $\bar{G}_s(\infty|\x_0) > 0$ does exist in the supercritical
regime for $s < 1$, and this solution determines the long-time
asymptotic behavior:
\begin{equation}  \label{eq:Gs_asympt_growth}
\bar{G}_s(t|\x_0) \approx \bar{G}_s(\infty|\x_0)  \qquad (t\to\infty).
\end{equation}

To gain some probabilistic insights, one can rewrite
Eq. (\ref{eq:Gs_infty}) as
\begin{equation}  \label{eq:Gs_infty1}
G_s(\infty|\x_0) = \pi(\x_0) + \qc \int\limits_{\Gammac} d\x \, \G^+(\x,\x_0) \, [G_s(\infty|\x)]^2,
\end{equation}
where
\begin{equation}
\pi(\x_0) = 1 - \qc \int\limits_{\Gammac} d\x \, \G^+(\x,\x_0).
\end{equation}
The integral equation (\ref{eq:Gs_infty1}) has a simple probabilistic
interpretation: in the steady-state regime, a particle started from
$\x_0$ is either absorbed on $\Gammaa$ with probability $\pi(\x_0)$,
or branches at a point $\x\in\Gammac$, creating two identical
independent offsprings.  The probability density of branching at
$\x\in\Gammac$ is precisely $-\partial_n \G^+(\x,\x_0)\bigr|_{\Gammac}
= \qc \G^+(\x,\x_0)\bigr|_{\Gammac}$.  This extension of the spread
harmonic measure density \cite{Grebenkov06b,Grebenkov15b} includes the
condition that the particle is not absorbed on $\Gammaa$.  In turn,
$\pi(\x_0)$ is known as splitting probability, i.e., the probability
of reacting first on $\Gammaa$.

\subsection{Criterion of existence}

To inspect the existence of a nonzero solution, one can reformulate
the problem in a more formal way in terms of operators.  In fact,
restricting $\x_0 \in \Gammac$ in Eq. (\ref{eq:Gs_infty}), one can
rewrite it as
\begin{equation}  \label{eq:Gs_infty3}
\bar{G}_s(\infty|\x_0) =  \qc \G^+ \bigl\{2\bar{G}_s(\infty|\x) - [\bar{G}_s(\infty|\x)]^2\bigr\},
\end{equation} 
where $\G^+$ is an integral operator with the kernel $\G^+(\x,\x_0)$,
acting on a suitable function on $\Gammac$.  While the Green's
function $\G^+(\x,\x_0)$ admits the spectral expansion
(\ref{eq:GLaplace_spectral}) over Laplacian eigenbasis, it is not
suitable for the analysis of the operator $\G^+$, which acts in the
functional space $L^2(\Gammac)$.  To overcome this limitation, we
consider the eigenpairs $\{\mu_k, v_k\}$ of the following Steklov
spectral problem \cite{Grebenkov26}
\begin{subequations}  \label{eq:vk_Steklov}
\begin{align}  \label{eq:vk_diff}
\Delta v_k & = 0 \quad \textrm{in}~\Omega, \\  \label{eq:vk_qc}
\partial_n v_k &= \mu_k v_k \quad \textrm{on}~\Gammac, \\  \label{eq:vk_qa}
\partial_n v_k &= - \qa v_k \quad \textrm{on}~\Gammaa, \\
\partial_n v_k & = 0 \quad \textrm{on}~\Gammar
\end{align}
\end{subequations}
(in the case $\qa = \infty$, the Robin condition (\ref{eq:vk_qa}) is
replaced by the Dirichlet condition $v_k = 0$ on $\Gammaa$).  The
peculiar feature of the Steklov problem is that the spectral parameter
$\mu_k$ stands in the boundary condition (\ref{eq:vk_qc}).  The
spectrum is known to be discrete \cite{Levitin}, with infinitely many
eigenvalues $\{\mu_k\}$, which are enumerated to form an increasing
sequence,
\begin{equation}
0 < \mu_0 < \mu_1 \leq \mu_2 \leq \cdots \nearrow +\infty ,
\end{equation}
whereas the restrictions of the associated eigenfunctions $\{v_k\}$
onto $\Gammac$ form a complete orthonormal basis of $L^2(\Gammac)$
such that
\begin{equation}
\int\limits_{\Gammac} d\x \, v_j(\x) \, v_k(\x) = \delta_{j,k} .
\end{equation}
The principal eigenvalue $\mu_0$ is simple (of multiplicity $1$),
whereas the associated eigenfunction $v_0(\x)$ does not change the
sign in $\overline{\Omega}$ and can thus be chosen to be positive due
to the Courant nodal domain theorem.
Extending the derivation from \cite{Grebenkov20}, one can show that
the kernel of the operator $\G^+$ admits the spectral expansion over
the eigenfunctions $\{v_k|_{\Gammac}\}$:
\begin{equation}  \label{eq:G_vk}
\G^+(\x,\x_0) = \sum\limits_{k=0}^\infty \frac{v_k(\x) v_k(\x_0)}{\mu_k + \qc}   \qquad (\x,\x_0\in\Gammac).
\end{equation}
Since the eigenvalues $\mu_k$ are positive and accumulate at infinity,
$\G$ is a compact self-adjoint operator.  Denoting by ${\mathcal I}$
the identity operator, one can rewrite Eq. (\ref{eq:Gs_infty3}) as
\begin{equation}  \label{eq:barGsinf_A}
\bar{G}_s(\infty|\x_0) = \A \bigl\{\bar{G}_s(\infty|\x)- [\bar{G}_s(\infty|\x)]^2\bigr\}  \quad (\x_0\in \Gammac),
\end{equation} 
where $\A = [{\mathcal I} - \qc \G^+]^{-1} \qc \G^+$.  Using the
spectral expansion (\ref{eq:G_vk}), one can see that the new operator
$\A$ is again a compact self-adjoint positive-definite operator acting
on a function $f$ as
\begin{equation}
[\A f](\x_0) = \qc \sum\limits_{k=0}^\infty  \frac{v_k(\x_0)}{\mu_k} \int\limits_{\Gammac} d\x \, v_k(\x) \, f(\x).
\end{equation}

The problem of finding a nontrivial solution of the integral equation
(\ref{eq:Gs_infty1}) is thus reformulated in terms of the operator
equation (\ref{eq:barGsinf_A}).  This is a standard problem in the
bifurcation theory of nonlinear equations.  While $\bar{G}_s \equiv 0$
is always a solution of this equation, the existence of another
solution depends on the largest eigenvalue of the operator $\A$, which
is equal to $\qc/\mu_0$: if $\qc/\mu_0 < 1$, $\bar{G}_s \equiv 0$ is
the unique solution, whereas for $\qc/\mu_0 > 1$, there exists a
nontrivial solution.  This criterion naturally appears by considering
$\bar{G}_s^2$ as a perturbation and inspecting the linearized problem
$\bar{G}_s \approx \A \bar{G}_s$, in which the repeated application of
$\A$ would either contract the solution to $0$ (if $\qc/\mu_0 < 1$),
or blow it up to infinity (if $\qc/\mu_0 > 1$).  Since $0\leq
\bar{G}_s \leq 1$, the correction term $\bar{G}_s^2$ cannot alter the
contraction in the first case, yielding $\bar{G}_s\equiv 0$ as the
unique solution.  In turn, the correction term amends blowing up in
the second case, ensuring the convergence to a nontrivial solution.  A
rigorous proof of the above statements is beyond the scope of this
work.

In the marginal case $\qc/\mu_0 = 1$, one can show that there is no
nontrivial solution.  For this purpose, one can multiply
Eq. (\ref{eq:barGsinf_A}) by $v_0(\x_0)$ and integrate over $\Gammac$
that yields, due to the self-adjoint character of the operator $\A$:
\begin{align} \nonumber
\biggl(1 - \frac{\qc}{\mu_0}\biggr) & \int\limits_{\Gammac} d\x_0 \, v_0(\x_0) \bar{G}_s(\infty|\x_0) \\
& = -\frac{\qc}{\mu_0} \int\limits_{\Gammac} d\x \, v_0(\x) [\bar{G}_s(\infty|\x)]^2  .
\end{align}   
Since both $\bar{G}_s(\infty|\x_0)$ and $v_0(\x_0)$ are nonnegative,
this equality is only possible if $\qc/\mu_0 > 1$.  In the marginal
case, the left-hand side is zero, implying $\bar{G}_s(\infty|\x)
\equiv 0$.

As discussed in \cite{Grebenkov26}, the smallest eigenvalue $\mu_0$,
which depends on $\qa$, separates the subcritical ($\qc < \mu_0$) and
supercritical ($\qc > \mu_0$) regimes.  In fact, the PDE problems
(\ref{eq:vk_Steklov}) and (\ref{eq:Laplace}) become identical when
$\lambda_0 = 0$ and $\mu_0 = \qc$, i.e., $\mu_0$ determines the
critical catalytic rate $\qc$, at which the system reaches a nonzero
steady-state limit (the critical regime).  In turn, the variational
principle and monotonicity imply that $\lambda_0 > 0$ for $\qc <
\mu_0$ (the subcritical regime) and $\lambda_0 < 0$ for $\qc >
\mu_0$ (the supercritical regime).  Note that, in \cite{Grebenkov26},
a slightly different Steklov problem was used, in which the spectral
parameter was set on $\Gammaa$, whereas $\qc$ was fixed on $\Gammac$.
In that case, the smallest eigenvalue determined, for a given $\qc$,
the critical absorption rate $\qa$ that separates the subcritical and
supercritical regimes.

The asymptotic approach to the limiting value $\bar{G}_0(\infty|R)$ is
illustrated on Fig. \ref{fig:Q0_asympt2}(c).

\subsection{Long-time behavior of the distribution}

The above asymptotic analysis can be easily transferred to the
probabilities $Q_k(t|\x_0)$ that are determined by the generating
function $G_s(t|\x_0)$ via Eq. (\ref{eq:Qn_def}).  Since the behavior
of $Q_0(t|\x_0) = G_0(t|\x_0)$ was already discussed in the previous
subsections, we sketch the asymptotic results for other probabilities
with $k > 0$ for three regimes:

(i) In the subcritical regime, Eq. (\ref{eq:barGs_asympt_ext})
immediately implies the same asymptotic behavior,
\begin{equation}  
Q_k(t|\x_0) \simeq q_k \, u_0^-(\x_0) \, C_0^- \, e^{-Dt\lambda_0^-}  \quad (t\to \infty),
\end{equation}
where the prefactor $q_k$ is obtained by differentiating the amplitude
$g_s$ from Eq. (\ref{eq:barGs_asympt_a}):
\begin{align} 
q_k & = - \frac{1}{k!} \lim\limits_{s\to 0} \partial_s^k g_s  \\  \nonumber
& = \delta_{k,1} + \frac{\qc D}{C_0^-} \int\limits_{\Gammac} d\x \, u_0^-(\x) \int\limits_0^\infty dt' \, e^{Dt'\lambda_0^-} H_k(t'|\x),
\end{align}
with $H_k(t|\x)$ given by Eq. (\ref{eq:Hn_def}).

(ii) In the critical regime, the leading-order term of
$\bar{G}_s(t|\x_0)$ in Eq. (\ref{eq:barGs_asympt_crit}) does not
depend on $s$ and thus disappears after differentiation.  One can
therefore expect that the probabilities $Q_k(t|\x_0)$ decay faster
than $\bar{Q}_0(t|\x_0) \propto 1/t$.  We conjecture the following
asymptotic behavior for $k = 1,2,\ldots$:
\begin{equation}  \label{eq:Qn_asympt_crit}
Q_k(t|\x_0) \simeq q_k\, u_0^-(\x_0)\, t^{-2}  \quad (t\to\infty),
\end{equation}
with some prefactors $q_k$.  

(iii) In the supercritical regime, we showed that $\bar{G}_s(t|\x_0)$
approaches a nontrivial limit, and so does the probability
$Q_0(t|\x_0)$.  As this limit does not depend on $s$, the
probabilities $Q_k(t|\x_0)$ should vanish for any $k > 0$.  The
asymptotic analysis of their decay presents an interesting open
problem.

\section{Moments of the population size}
\label{sec:moments}

The generating function $G_s(t|\x_0)$ also determines the moments of
the population size $\N(t)$ of all positive-integer orders $k =
1,2,\ldots$
\begin{equation}  \label{eq:Nk_def}
N_k(t|\x_0) = \E_{\x_0}\{ [\N(t)]^k\} = \lim\limits_{s\to 1} \biggl[(s\partial_s)^k G_s(t|\x_0)\biggr].
\end{equation}
For instance, the first derivative with respect to $s$ of
Eqs. (\ref{eq:Gs_PDE}), evaluated at $s=1$, yields the diffusion
problem for the mean population size $N_1(t|\x_0)$:
\begin{subequations}  \label{eq:Nt_system}
\begin{align}
\partial_t N_1(t|\x_0) - D \Delta N_1(t|\x_0) & = 0 \quad \textrm{in}~\Omega, \\  \label{eq:Nt_qc}
\partial_n N_1(t|\x_0) - \qc N_1(t|\x_0) & = 0 \quad \textrm{on}~ \Gammac, \\  \label{eq:Nt_qa}
\partial_n N_1(t|\x_0) + \qa N_1(t|\x_0) & = 0 \quad \textrm{on}~ \Gammaa, \\
\partial_n N_1(t|\x_0) & = 0 \quad \textrm{on}~ \Gammar, \\
N_1(t=0|\x_0) & = 1 .
\end{align}
\end{subequations}
Despite the branching mechanism, this PDE is {\it linear}.  In fact,
we retrieved the usual description of the survival probability, except
that the surface reactivity is {\it negative} on the catalytic region.
This problem was thoroughly investigated in \cite{Grebenkov26}.  While
the absorption events on $\Gammaa$ reduce the population size and thus
yield the positive diffusive flux from the bulk to the surface,
$-\partial_n N_1(t|\x_0)\bigr|_{\Gammaa} = \qa N_1(t|\x_0) > 0$, the
branching events increase the population size on $\Gammac$, resulting
in the opposite direction of the diffusive flux: $-\partial_n
N_1(t|\x_0)\bigr|_{\Gammac} = -\qc N_1(t|\x_0) < 0$.  Integrating the
propagator $P^-(\x,t|\x_0)$ defined by Eqs. (\ref{eq:propagator}), one
finds
\begin{equation}  \label{eq:N1_P}
N_1(t|\x_0) = \int\limits_{\Omega} d\x \, P^-(\x,t|\x_0) = S^-(t|\x_0),
\end{equation}
which also gives a simple interpretation of the function $S^-(t|\x_0)$
used in Sec. \ref{sec:Gs_asympt}.

\begin{figure}
\begin{center}
\includegraphics[width=0.99\columnwidth]{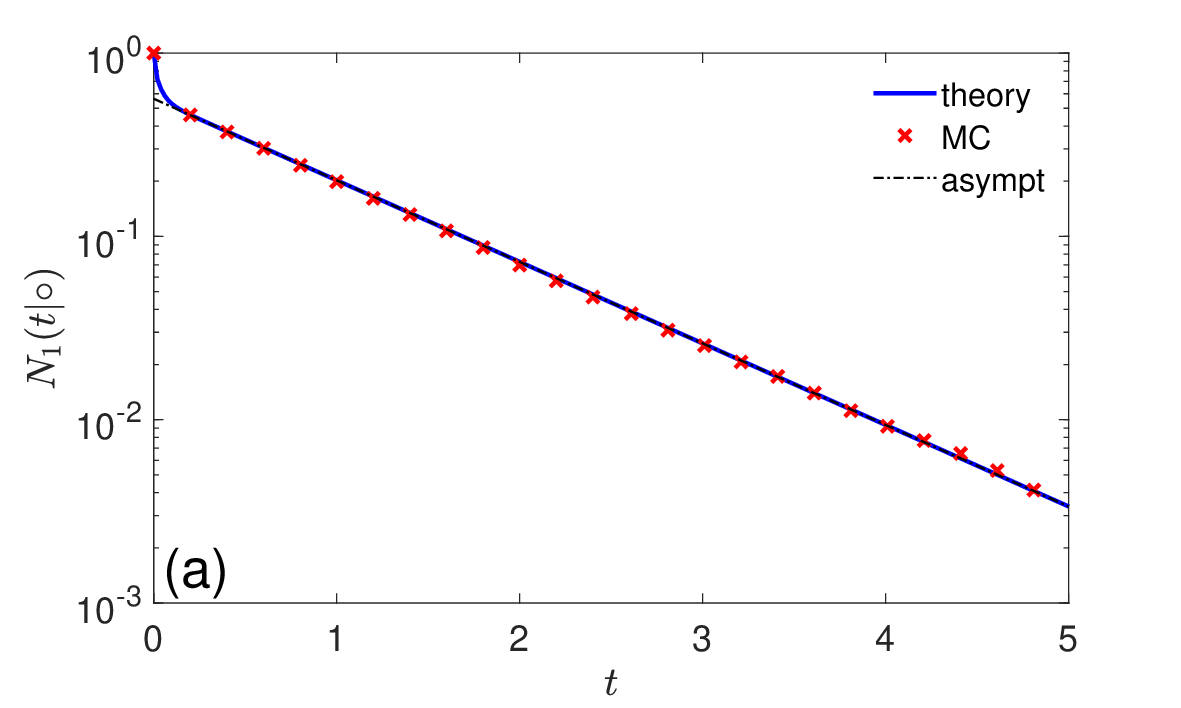} % {N1_qc09_new.eps}
\includegraphics[width=0.99\columnwidth]{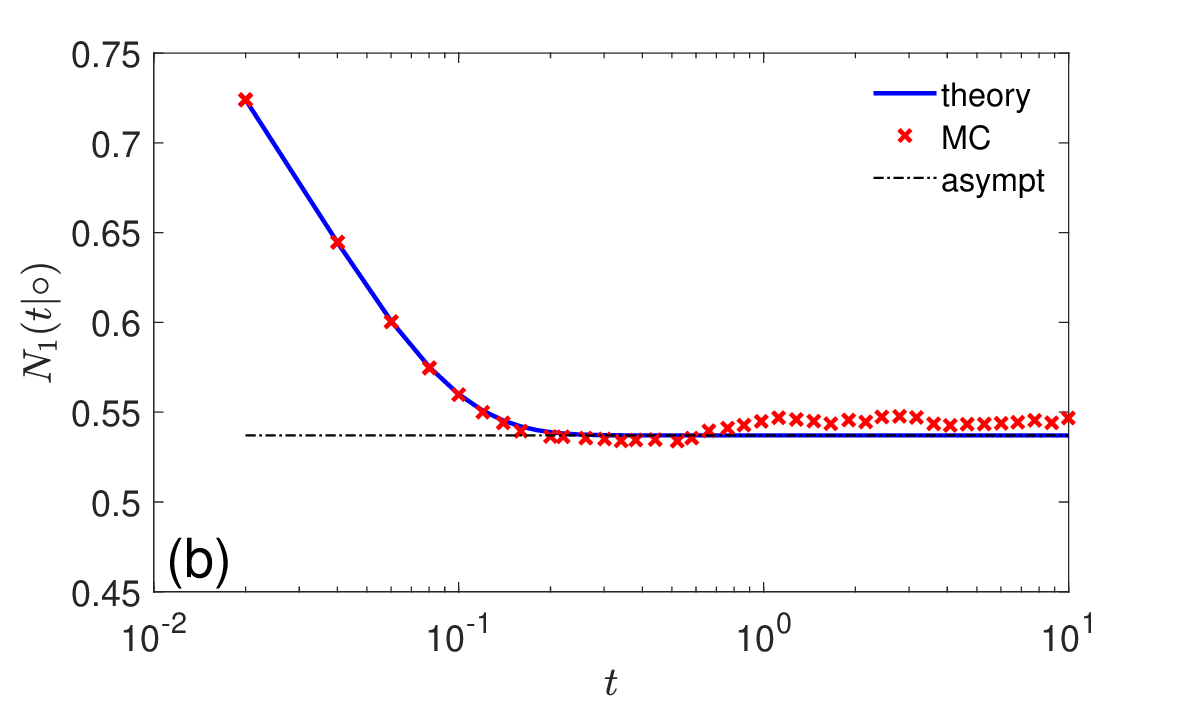} % {N1_qc10_new.eps}
\includegraphics[width=0.99\columnwidth]{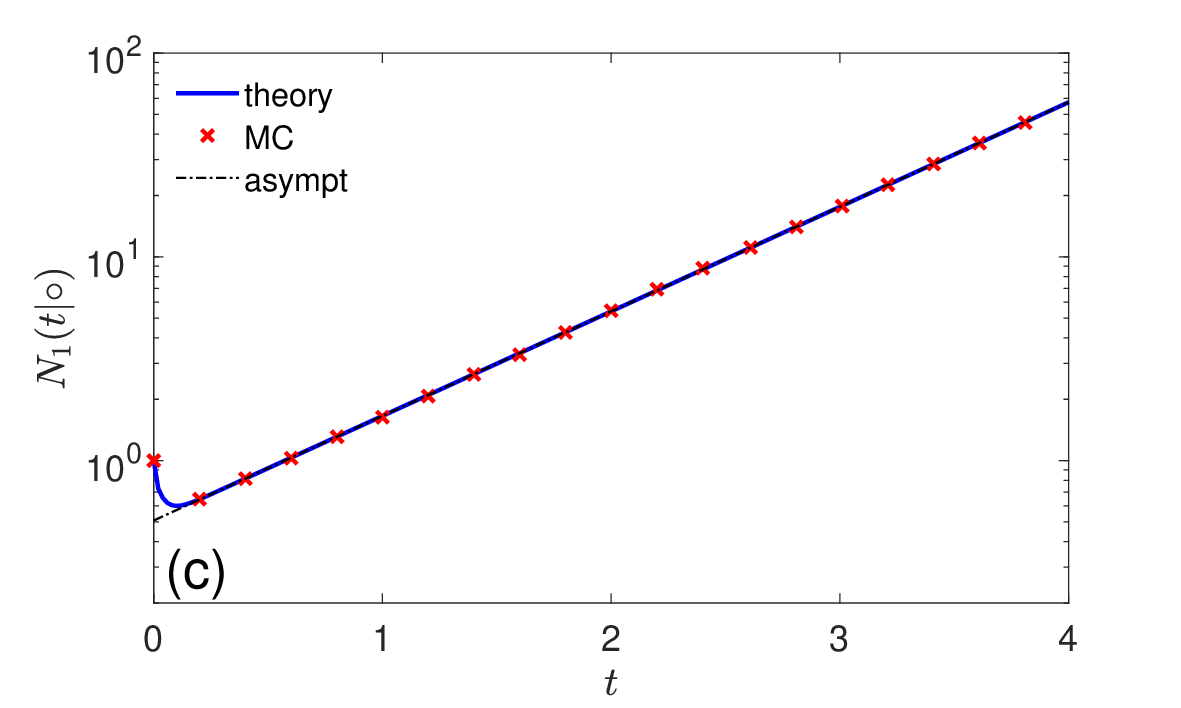} % {N1_qc11.eps}
\end{center}
\caption{
Mean population size $N_1(t|\circ)$ for the circular annulus with $R =
0.1$, $L = 1$, $D = 1$, $\qa = \infty$, uniform starting point in the
bulk.  Three panels illustrate three regimes: {\bf (a)} subcritical
($q_{\c} = 0.9\, \qccrit \approx 3.91$); {\bf (b)} critical ($q_{\c} =
\qccrit \approx 4.34$); and {\bf (c)} supercritical ($q_{\c} = 1.1\,
\qccrit \approx 4.78$), with $\qccrit = 1/(R\ln(L/R))$.  Solid blue line
presents the exact result obtained via the inverse Laplace transform
of Eq. (\ref{eq:Np1_annulus}); crosses show the empirical mean
obtained over $10^6$ Monte Carlo simulations (Appendix \ref{sec:MC}),
whereas dashed line indicates the asymptotic relation
(\ref{eq:N1_asympt}).}
\label{fig:Nt_mean}
% [Nt] = A_Yilin4_simu12(2);
% [Nt] = A_Yilin4_simu12(1);
% [Nt] = A_Yilin4_simu12(3);    % we used 1e5 particles for qc = 1.1
\end{figure}

Figure \ref{fig:Nt_mean} illustrates the behavior of the mean
population size $N_1(t|\circ)$ inside the circular annulus $\Omega =
\{ \x\in\R^2 ~:~ R < |\x| < L\}$ for the initial particle with a
uniformly distributed starting point in the bulk (as indicated by
$\circ$ instead of $\x_0$):  
\begin{equation}
N_1(t|\circ) = \frac{1}{|\Omega|} \int\limits_{\Omega} d\x_0 \, N_1(t|\x_0).
\end{equation}
The empirical mean from Monte Carlo simulations is in excellent
agreement with the exact solution, which is obtained via the inverse
Laplace transform of Eq. (\ref{eq:Np1_annulus}) by using a Talbot
algorithm \cite{Talbot79}.  Depending on the value of the catalytic
rate $\qc$, one observes an exponential decrease ($\qc < \qccrit$), a
steady-state ($\qc = \qccrit$), and an exponential growth ($\qc >
\qccrit$) of the mean population size, where $\qccrit = \mu_0(\infty)$
is the principal eigenvalue of the Steklov problem
(\ref{eq:vk_Steklov}), with $\qa = \infty$.

In turn, the $k$-th order derivative of Eqs. (\ref{eq:Gs_PDE}) with
respect to $s$ yields
\begin{subequations}  \label{eq:Ntk_system}
\begin{align}
\partial_t N_k(t|\x_0) - D \Delta N_k(t|\x_0) & = 0 \quad \textrm{in}~\Omega, \\  \label{eq:Ntk_qc}
\partial_n N_k(t|\x_0) - \qc N_k(t|\x_0) & = \qc  F_k(t|\x_0) \quad \textrm{on}~ \Gammac, \\ 
\partial_n N_k(t|\x_0) + \qa N_k(t|\x_0) & = 0 \quad \textrm{on}~ \Gammaa, \\
\partial_n N_k(t|\x_0) & = 0 \quad \textrm{on}~ \Gammar, \\
N_k(t=0|\x_0) & = 1 ,
\end{align}
\end{subequations}
where we used Eqs. (\ref{eq:Nt_system}), and
\begin{equation}  \label{eq:Fk_def0}
F_k(t|\x_0) = -2N_k(t|\x_0) + \lim\limits_{s\to 1} (s \partial_s)^k G_s^2(t|\x_0).
\end{equation}
Here we included the term $-2N_k(t|\x_0)$ to get the negative $\qc$ in
the left-hand side of Eq. (\ref{eq:Ntk_qc}).  Using the $k$-th order
Leibniz rule to evaluate the action of the operator $(s\partial_s)^k$
on the product of two functions (here, the square of $G_s(t|\x_0)$),
one gets an explicit form for any $k \geq 2$:
\begin{equation}  \label{eq:Fk_def}
F_k(t|\x_0) = \sum\limits_{j=1}^{k-1} \binom{k}{j} N_j(t|\x_0) \, N_{k-j}(t|\x_0).
\end{equation}
For instance, one has
\begin{subequations}  \label{eq:fk_def}
\begin{align}  \label{eq:f2_def}
F_2(t|\x_0) & = 2[N_1(t|\x_0)]^2,  \\
F_3(t|\x_0) & = 6 N_2(t|\x_0) N_1(t|\x_0), \\
F_4(t|\x_0) & = 6 [N_2(t|\x_0)]^2 + 8 N_1(t|\x_0) N_3(t|\x_0).
\end{align}
\end{subequations} 

While the mean population size $N_1(t|\x_0)$ satisfies the {\it
homogeneous} Robin boundary condition (\ref{eq:Nt_qc}) on $\Gammac$,
the second and higher-order moments involve the inhomogeneous Robin
condition (\ref{eq:Ntk_qc}), with an additional ``source'' term $\qc
F_k(t|\x_0)$ that is expressed through the lower-order moments.  The
crucial point is that the initial-value problem (\ref{eq:Ntk_system})
is {\it linear} in $N_k(t|\x_0)$, whereas the nonlinearity of the
boundary condition (\ref{eq:Gs_qc}) for the generating function is
reflected in the nonlinear structure of the source term $F_k(t|\x_0)$.

The linear structure of the PDE allows one to expression its solution
in terms of the propagator $P^-(\x,t|\x_0)$.  Following the standard
technique (see Appendix \ref{sec:technical}), we get
\begin{align}  \label{eq:Nk_fk}
N_k(t|\x_0) & = S^-(t|\x_0)  \\  \nonumber
& + \qc D \int\limits_{\Gammac} d\x \int\limits_0^t dt' \, P^-(\x,t'|\x_0) \, F_k(t-t'|\x).
\end{align}
Since $F_k(t|\x)$ depends on the lower-order moments, one can evaluate
the moments $N_k(t|\x)$ iteratively, if the propagator $P^-(\x,t|\x_0)$
is known (see Appendix \ref{sec:numerics} for details).

We complete this section by inspecting the long-time behavior.
First, we write the spectral expansion for the mean population size:
\begin{equation}  \label{eq:N1_spectral}
N_1(t|\x_0) = \sum\limits_{k=0}^\infty C_k^- \, u_k^-(\x_0) \, e^{-Dt\lambda_k^-} , 
\end{equation}
where the coefficients $C_k^-$ were defined in Eq. (\ref{eq:Ck_def}).
One sees that
\begin{equation}  \label{eq:N1_asympt}
N_1(t|\x_0) \simeq C_0^- u_0^-(\x_0) e^{-Dt\lambda_0^-}  \quad (t\to \infty).  
\end{equation}
Next, according to Eq. (\ref{eq:f2_def}), this expansion determines
the source term $F_2(t|\x_0)$ for the second moment, which determines
the source term $F_3(t|\x_0)$ for the third moment, and so.  As the
asymptotic behavior depends on the sign of the principal eigenvalue
$\lambda_0^-$, we distinguish three regimes.

\subsection{Subcritical regime}

When $\lambda_0^- > 0$, one can employ the same analysis as in
Sec. \ref{eq:barGs_ext} to show that 
\begin{equation}  \label{eq:Nk_asympt_ext}
N_k(t|\x_0) \simeq n_k u_0^-(\x_0) C_0^- e^{-Dt\lambda_0^-} \quad (t\to\infty),
\end{equation}
with a dimensionless amplitude $n_k$.  Since $F_k(t|\x_0)$ is
expressed via Eq. (\ref{eq:Fk_def}) in terms of products of
lower-order moments, one can resort to the induction argument, i.e.,
one can assume that Eq. (\ref{eq:Nk_asympt_ext}) holds for
$j=1,2,\ldots,k-1$ and then check it for $k$.  This assumption implies
the long-time behavior
\begin{equation}  \label{eq:Fk_asympt}
F_k(t|\x_0) \simeq (u_0^-(\x_0) C_0^-)^2 e^{-2Dt\lambda_0^-} \sum\limits_{j=1}^{k-1} \binom{k}{j} n_j n_{k-j} ,
\end{equation}
so that the Laplace transform of $F_k(t|\x_0)$ does not have a pole at
$p_0 = -D\lambda_0^-$.  As a consequence, the long-time asymptotic
behavior of the right-hand side of Eq. (\ref{eq:Nk_fk}) is
\begin{align*}
& N_k(t|\x_0) \simeq  C_0^- u_0^-(\x_0) e^{-Dt\lambda_0^-}   \\  \nonumber
& + \qc D u_0^-(\x_0) e^{-Dt\lambda_0^-} \int\limits_{\Gammac} d\x\, u_0^-(\x) \int\limits_0^\infty dt' \, e^{Dt'\lambda_0^-} \, F_k(t'|\x),
\end{align*}
that is reduced to Eq. (\ref{eq:Nk_asympt_ext}) by setting
\begin{equation}  \label{eq:nk_formal}
n_k = 1 + \frac{\qc D }{C_0^-} \int\limits_{\Gammac} d\x\, u_0^-(\x) \int\limits_0^\infty dt' \, e^{Dt'\lambda_0^-} \, F_k(t'|\x).
\end{equation}

As in Sec. \ref{eq:barGs_ext}, the inconvenience of this
representation is that one needs to know $F_k(t'|\x)$ for all $t' >
0$.  If the long-time relation (\ref{eq:Fk_asympt}) could be used for
the whole range of times, the above expression (\ref{eq:nk_formal})
would be significantly simplified:
\begin{equation}  \label{eq:nk_ext}
n_k \approx 1 + \frac{q_0 \qc}{\lambda_0^-} \sum\limits_{j=1}^{k-1} \binom{k}{j} n_j n_{k-j} ,
\end{equation}
with $q_0$ defined by Eq. (\ref{eq:q0}).  Using $n_1 = 1$ according to
Eq. (\ref{eq:N1_asympt}), one can iteratively compute these
amplitudes; for instance, one has $n_2 \approx 1 + 2q_0
\qc/\lambda_0^-$.  

\begin{figure}[!ht]
\begin{center}
\includegraphics[width=0.99\columnwidth]{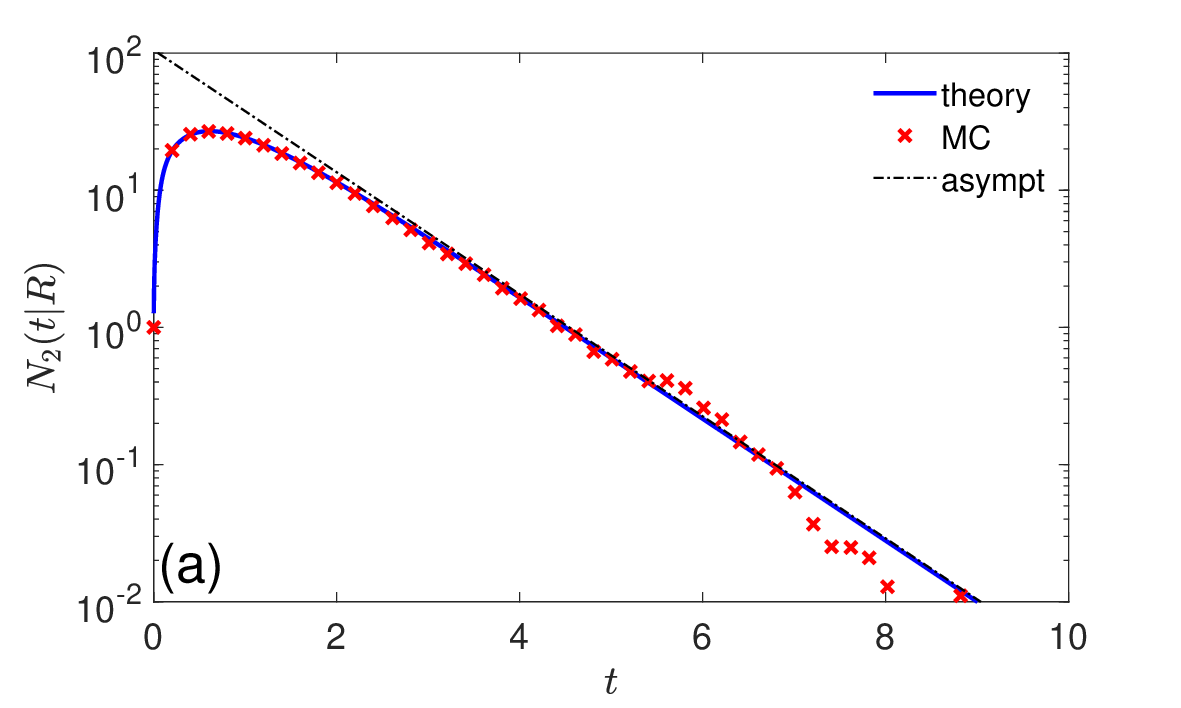} % {N2_qc09_new.eps}
\includegraphics[width=0.99\columnwidth]{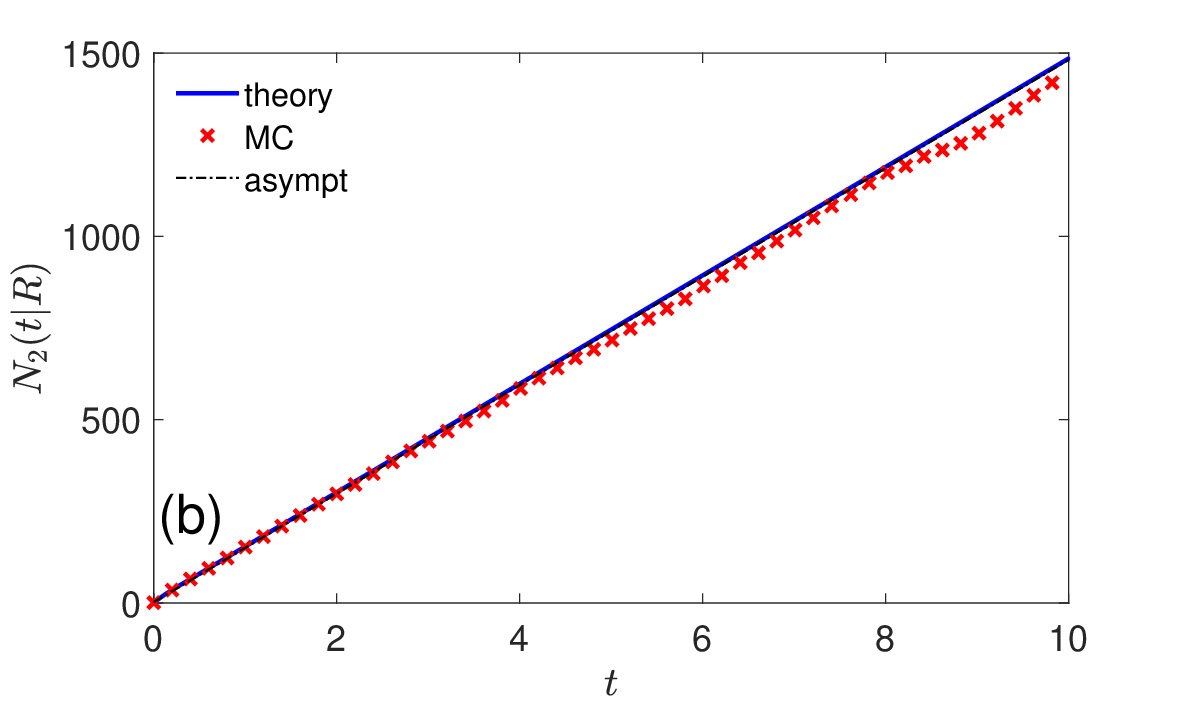} % {N2_qc10_new.eps}
\includegraphics[width=0.99\columnwidth]{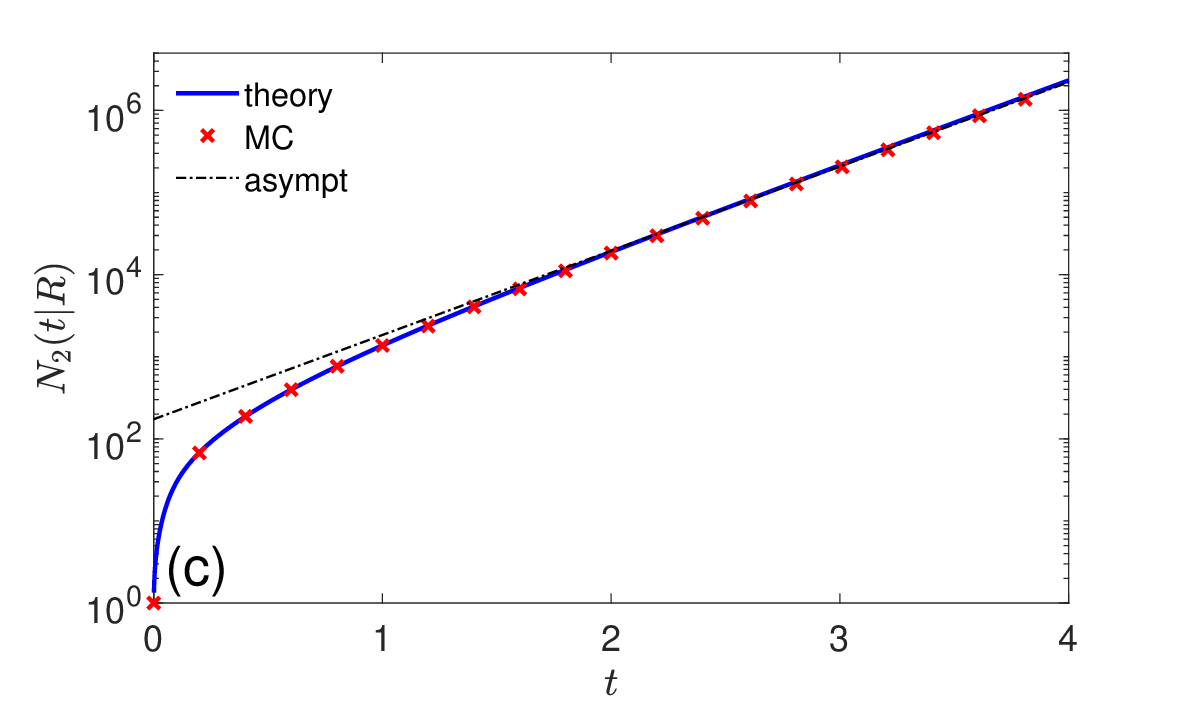} % {N2_qc11.eps}
\end{center}
\caption{ 
The second moment of the population size, $N_2(t|R)$, for the circular
annulus with $R = 0.1$, $L = 1$, $D = 1$, $\qa = \infty$, and the
starting point $\x_0$ on the catalytic region.  Three panels
illustrate three regimes: {\bf (a)} subcritical ($\qc = 0.9\, \qccrit
\approx 3.91$); {\bf (b)} critical ($\qc = \qccrit \approx 4.34$); and
{\bf (c)} supercritical ($\qc = 1.1\, \qccrit \approx 4.78$), with
$\qccrit = 1/(R\ln(L/R))$.  Solid blue line presents the exact result
obtained via a numerical solution of the integral equation
(\ref{eq:Nk_fk}) as described in Appendix \ref{sec:numerics}; crosses
show the empirical mean of $\N^2(t)$ obtained over $10^6$ random
realizations (see Appendix \ref{sec:MC}); dashed line indicates the
long-time asymptotic behavior: Eqs. (\ref{eq:Nk_asympt_ext},
\ref{eq:nk_ext}) for panel (a), Eq. (\ref{eq:N2_asympt_steady}) for
panel (b), and Eqs. (\ref{eq:Ntk_asympt_growth}, \ref{eq:Ak_growth},
\ref{eq:nk_growth}) for panel (c).}
\label{fig:Nt_mean2}
% [Nt] = A_Yilin4_simu13(2);
% [Nt] = A_Yilin4_simu13(1);
% [Nt] = A_Yilin4_simu13(3);  % I still used 1e5 particles
\end{figure}

Figure \ref{fig:Nt_mean2}(a) illustrates the exponential decay of the
second moment $N_2(t|R)$ when the starting point is located on the
catalytic region of the circular annulus.  The asymptotic relation
(\ref{eq:Nk_asympt_ext}) with the approximate form (\ref{eq:nk_ext})
for $n_2$ provides an accurate approximation.  Since $N_2(t|R)$
rapidly decreases, its accurate estimation from Monte Carlo
simulations requires too many realizations (see dispersion of crosses
at long times).

One practical consequence of the asymptotic relation
(\ref{eq:Nk_asympt_ext}) is that the coefficient of variation, i.e.,
the ratio between the standard deviation and the mean,
\begin{equation}
\gamma(t|\x_0) = \frac{\sqrt{N_2(t|\x_0) - N_1^2(t|\x_0)}}{N_1(t|\x_0)}
\propto e^{Dt\lambda_0^-/2} ,
\end{equation}
diverges as $t\to\infty$.  In other words, fluctuations become more
and more relevant in the subcritical regime so that rare realizations
of the stochastic process $\N(t)$ can significantly affect its
moments, as illustrated on Fig. \ref{fig:Nt_traj}(a).

\subsection{Critical regime}

In the critical regime ($\lambda_0^- = 0$), both $P^-(\x,t|\x_0)$ and
$F_2(t|\x_0)$ reach nonzero steady-state limits so that the integral
over $t'$ in Eq. (\ref{eq:Nk_fk}) yields $t$ in the leading order.
More explicitly, we have
\begin{equation}  \label{eq:N2_asympt_steady}
N_2(t|\x_0) \simeq u_0^-(\x_0) C_0^- \bigl(1 + 2Dt \,\qc q_0 \bigr),
\end{equation}
with exponentially decaying correction terms.  In the Laplace domain,
this behavior means that the product of two functions with a simple
pole $p_0 = 0$ yields a function with a double pole $p_0 = 0$.
Repeating this procedure iteratively, one gets
\begin{equation}
N_k(t|\x_0) \propto t^{k-1}  \quad (t\to\infty).
\end{equation}
In this case, the fluctuations are also dominant, and the coefficient
of variation exhibits a square-root growth with time:
$\gamma(t|\x_0)\propto \sqrt{t}$.  

Figure \ref{fig:Nt_mean2}(b) illustrates the remarkable accuracy of
the asymptotic relation (\ref{eq:N2_asympt_steady}) in the case of a
circular annulus (the dashed line is barely distinguishable from the
solid line).  In turn, deviations of Monte Carlo estimates at long
times highlight the difficulty in performing accurate simulations in
the critical regime.

\subsection{Supercritical regime}

In the supercritical regime ($\lambda_0^- < 0$), the Laplace-transform
argument is still applicable but gives a different result.  In fact,
the Laplace-transformed propagator $\tilde{P}^-(\x,p|\x_0)$ still has
the largest simple pole $p_0 = - D\lambda_0^-$, which is now positive.
In turn, the product of lower-order moments in $F_k(t|\x_0)$ results
in the largest pole $-kD\lambda_0^-$, which is greater than $p_0$, and
thus it controls the long-time behavior.  However, since the
asymptotic behavior of the propagator $P^-(\x,t|\x_0)$ is not the
dominant factor, one loses the explicit dependence on the starting
point $\x_0$ via the principal eigenfunction $u_0^-(\x_0)$.  In turn,
one can still employ the induction argument, i.e., to assume that
\begin{equation}  \label{eq:Ntk_asympt_growth}
N_{k'}(t|\x_0) \simeq n_{k'}(\x_0) e^{k' Dt|\lambda_0^-|}  \quad (t\to \infty)
\end{equation}
holds for $k' = 1,2,\ldots,k-1$ and then check it for $k$, where
$n_{k'}(\x_0)$ are (yet) unknown space-dependent amplitudes.  The
relation (\ref{eq:Ntk_asympt_growth}) implies $F_k(t|\x_0) \propto
e^{kDt|\lambda_0^-|}$ and this function determines the asymptotic
behavior of $N_k(t|\x_0)$.  Using the Laplace transform and the above
argument for the poles, we get the long-time behavior
\begin{align} \nonumber
N_k(t|\x_0) & \simeq \qc D e^{kDt|\lambda_0^-|} \int\limits_{\Gammac} d\x \, 
\biggl(\sum\limits_{j=1}^{k-1} \binom{k}{j} n_j(\x) n_{k-j}(\x) \biggr) \\
& \times \int\limits_0^\infty dt' \, e^{-kDt'|\lambda_0^-|} \, P^-(\x,t'|\x_0).
\end{align}
The dependence on the starting point $\x_0$ is thus formally given as
\begin{align}  \nonumber
n_k(\x_0) & \simeq \qc D \int\limits_{\Gammac} d\x \, 
\biggl(\sum\limits_{j=1}^{k-1} \binom{k}{j} n_j(\x) n_{k-j}(\x) \biggr) \\
& \times \int\limits_0^\infty dt' \, e^{-kDt'|\lambda_0^-|} \, P^-(\x,t'|\x_0).
\end{align}
If the propagator is known, one can determine $n_k(\x_0)$ iteratively,
starting from $n_1(\x_0) = C_0^{-} u_0^-(\x_0)$.

Using the spectral expansion (\ref{eq:p_spectral}), one
formally gets
\begin{equation}
\int\limits_0^\infty dt' \, e^{-kDt'|\lambda_0^-|} \, P^-(\x,t'|\x_0) = \sum\limits_{j=0}^\infty 
\frac{u_j^-(\x) u_j^-(\x_0)}{D(\lambda_j^- + k|\lambda_0^-|)} \,,
\end{equation}
so that the dependence on $\x_0$ involves all eigenfunctions
$u_j^-(\x_0)$.  However, if all eigenfunctions except $u_0^-(\x)$
could be neglected, the spatial dependence of $n_k(\x)$ would be
determined by the principal one,
\begin{equation}  \label{eq:Ak_growth}
n_k(\x) \approx n_k \, C_0^- \, u_0^-(\x) ,
\end{equation}
in which case the dimensionless coefficients $n_k$ would satisfy
\begin{equation}  \label{eq:nk_growth}
n_k \approx \frac{q_0 \qc}{|\lambda_0^-|(k-1)} \sum\limits_{j=1}^{k-1} \binom{k}{j} n_j n_{k-j} ,
\end{equation}
with $q_0$ being defined by Eq. (\ref{eq:q0}).  Using $n_1 = 1$
according to Eq. (\ref{eq:N1_asympt}), one can iteratively compute
these amplitudes; for instance, one has $n_2 \approx 2q_0
\qc/|\lambda_0^-|$.  Quite surprisingly, this rough approximation
turns out to be rather accurate for a circular annulus, see
Fig. \ref{fig:Nt_mean2}(c).

The asymptotic behavior (\ref{eq:Ntk_asympt_growth}) has an important
consequence.  In this regime, the coefficient of variation
$\gamma(t|\x_0)$ reaches a constant limit.  More generally, any
higher-order moment $N_k(t|\x_0)$ after rescaling by $[N_1(t|\x_0)]^k$
reaches a constant limit.  This indicates that the rescaled population
size, $\N(t)/N_1(t|\x_0)$, is expected to achieve a stationary
distribution at long times.  Once this distribution is found, the
long-time behavior of the system is fully described by the mean
population size $N_1(t|\x_0)$.

\section{Population extinction time}
\label{sec:extinction}

The main focus of the paper was on the stochastic dynamics of the
population size $\N(t)$.  A complementary insight onto this dynamics
can be achieved through first-passage times
\cite{Redner,Metzler,Grebenkov}.  Among various temporal aspects, the
population extinction time $\T_0$ plays the crucial role in many
applications as a proxy to quantify sustainability and robustness of
the considered diffusion-reaction system.  This is the first time
instance when the number of particles drops to $0$.  After reaching
this state, the system cannot produce any particle, so that $\N(t)$
remains zero after $\T_0$.  As a consequence, the probabilistic event
$\T_0 < t$ implies $\N(t) = 0$ so that
\begin{equation}
Q_0(t|\x_0) = \P_{\x_0}\{ \T_0 < t\} .
\end{equation}
In other words, the probability $Q_0(t|\x_0)$ of having no particle at
time $t$ is actually the cumulative distribution function of the
extinction time $\T_0$ \cite{Grebenkov26a}.  In turn, its probability
density is simply obtained by taking the time derivative: $J_0(t|\x_0)
= \partial_t Q_0(t|\x_0)$.  One sees that our former analysis of the
probability $Q_0(t|\x_0)$ gives an immediate access to the
distribution of the extinction time $\T_0$.  In this section, we
briefly discuss some basic properties of this distribution, while its
detailed study is beyond the scope of this paper.

We first recall that the probability $Q_0(t|\x_0)$ can be found by
solving the integral equation (\ref{eq:Q0_integral}) or
(\ref{eq:barQ0_P0}), or by solving the initial-value problem
(\ref{eq:Qn_PDE}) with $k = 0$.  The time derivative of
Eq. (\ref{eq:Q0_integral}) yields an integral equation for the
probability density:
\begin{align}  \nonumber
J_0(t|\x_0) & = J_\a(t|\x_0) + 2\qc D \int\limits_{\Gammac} d\x \int\limits_0^t dt' 
P^+(\x,t'|\x_0) \\
& \times Q_0(t-t'|\x) J_0(t-t'|\x),
\end{align}
where
\begin{equation}
J_{\a}(t|\x_0) = \qa D \int\limits_{\Gammaa} d\x\, P^+(\x,t|\x_0)
\end{equation}
is the flux of particles onto the absorbing boundary $\Gammaa$ at time
$t$.  In turn, the time derivative of Eq. (\ref{eq:barQ0_P0}) gives an
equivalent integral equation:
\begin{align}  \nonumber
J_0(t|\x_0) & = -\partial_t S(t|\x_0) - \qc D \int\limits_{\Gammac} d\x \int\limits_0^t dt' 
P(\x,t'|\x_0) \\  \label{eq:J0_int}
& \times \bigl[1 - 2Q_0(t-t'|\x)\bigr] J_0(t-t'|\x).
\end{align}

As discussed in Sec. \ref{sec:Gs_asympt}, the long-time asymptotic
behavior of the probability $Q_0(t|\x_0)$ is controlled by the
catalytic rate $\qc$: $1-Q_0(t|\x_0) \propto e^{-Dt\lambda_0^-}$ in
the subcritical regime, $1-Q_0(t|\x_0) \propto 1/t$ in the critical
regime, and $Q_0(t|\x_0) \to Q_0(\infty|\x_0) < 1$ in the
supercritical regime.  As a consequence, the mean and higher-order
moments of $\T_0$ are infinite in the critical and supercritical
regimes.  In turn, in the subcritical regime, one can find them as
\begin{equation}
\E_{\x_0}\{ \T_0^k\} = \int\limits_0^\infty dt \, t^k \, J_0(t|\x_0) = k \int\limits_0^\infty dt \, t^{k-1} \, (1 - Q_0(t|\x_0)).
\end{equation}

Let us briefly discuss the mean extinction time in the subcritical
regime.  Substituting $Q_0(t|\x_0)$ from Eq. (\ref{eq:Q0_integral}),
we get
\begin{align*}
& \E_{\x_0}\{ \T_0\} = \int\limits_0^\infty dt \, (1-Q_0(t|\x_0)) 
= \int\limits_0^\infty dt \, S^+(t|\x_0) \\
& + \qc D \int\limits_0^\infty dt 
\int\limits_{\Gammac} d\x \int\limits_0^t dt' P^+(\x,t'|\x_0)\,
\bigl[1- Q_0^2(t-t'|\x)\bigr] .
\end{align*}
The time integral in the second term can be interpreted as a sort of
the Laplace transform of a convolution evaluated at $p = 0$, yielding
\begin{align} \nonumber
\E_{\x_0}\{ \T_0\} & = T^+(\x_0) + \qc D \int\limits_{\Gammac} d\x \,
\tilde{P}^+(\x,0|\x_0) \\  \label{eq:T0_mean}
& \times  \int\limits_0^\infty dt \bigl[1- Q_0^2(t|\x)\bigr] ,
\end{align}
where
\begin{equation}
T^+(\x_0) = \int\limits_0^\infty dt \, S^+(t|\x_0)
\end{equation}
is the mean first-reaction time on either of regions $\Gammac$ or
$\Gammaa$.  While the representation (\ref{eq:T0_mean}) still depends
on the probability $Q_0(t|\x)$, the spatial dependence of the mean
$\E_{\x_0}\{ \T_0\}$ is now explicitly captured via the Green's
function $D\tilde{P}^+(\x,0|\x_0)$.  In addition, this relation
identified $T^+(\x_0)$ as a natural lower bound for the mean
extinction time.

\begin{figure}[!ht]
\begin{center}
\includegraphics[width=0.99\columnwidth]{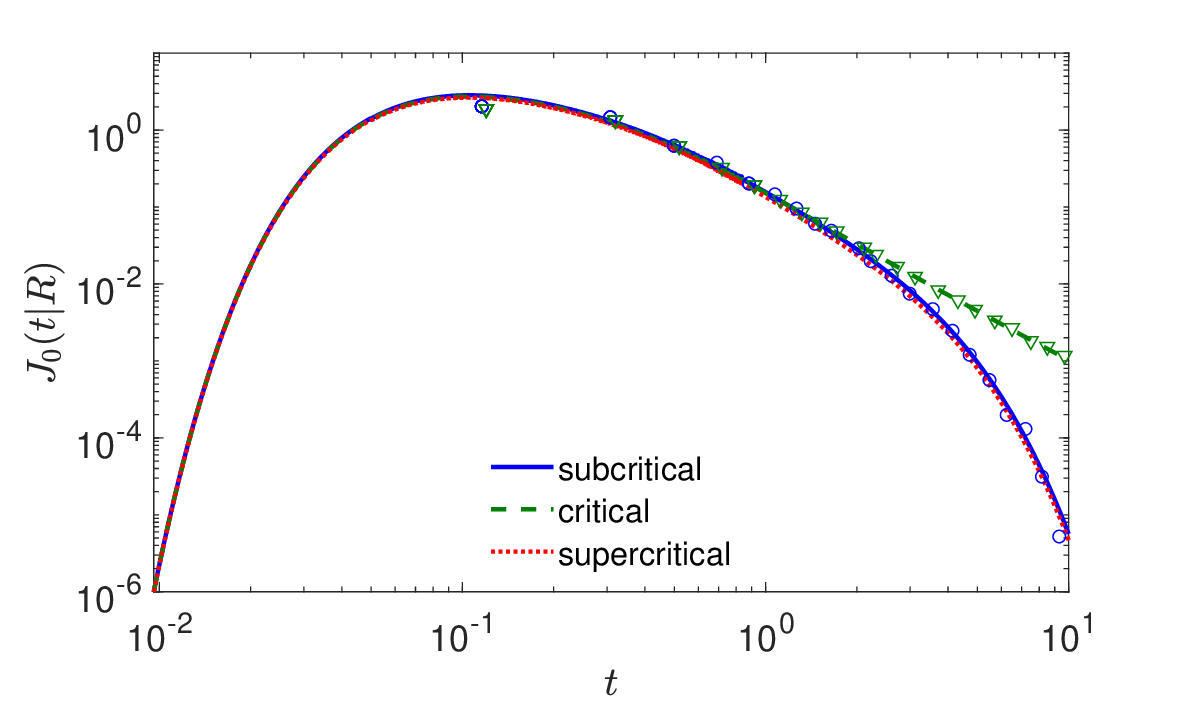} % {J0.eps}
\end{center}
\caption{
The probability density $J_0(t|R)$ of the population extinction time
$\T_0$ in the circular annulus with $R = 0.1$, $L = 1$, $D = 1$, $\qa
= \infty$, and the starting point $\x_0$ on the catalytic region.
Three curves correspond to three regimes: subcritical ($\qc = 0.9\,
\qccrit \approx 3.91$, solid blue line), critical ($\qc = \qccrit
\approx 4.34$, dashed green line), and supercritical ($\qc = 1.1\,
\qccrit \approx 4.78$, dotted red line), with $\qccrit =
1/(R\ln(L/R))$.  The function $J_0(t|R)$ was obtained at short times
by solving numerically the integral equation (\ref{eq:J0_int}); at
longer times, $J_0(t|R)$ was calculated directly from a
finite-difference approximation of the derivative of $Q_0(t|R)$.
Symbol present the empirical probability densities in the first two
regimes, which are estimated via Monte Carlo simulations from $10^6$
random realizations (see Appendix
\ref{sec:MC}). }
\label{fig:J0}
% [J1,Q1,J2,Q2,J3,Q3,t] = A_Yilin4_simu14c();
% For the circular annulus, we get
% $T^+(r_0) = (L^2 - r_0^2)/(4D) - A\ln(L/r)$, with $A = (L^2-R^2 + 2R/\qc)/(4D(\ln (L/R) + 1/(\qc R)))
\end{figure}

Figure \ref{fig:J0} shows the probability density $J_0(t|\x_0)$ for
the population dynamics in the circular annulus when the starting
point is located on the catalytic region at $|\x_0| = R$.  Three
curves represent three regimes corresponding to $\qc = 0.9\qccrit$,
$\qc = \qccrit$, and $\qc = 1.1 \qccrit$.  At short times, all three
curves are almost identical because the extinction of the population
requires crossing the domain from $\Gammac$ to $\Gammaa$ and is thus
highly unlikely.  In turn, the autocatalytic dynamics start to play
the dominant role at longer times; in particular, one can see a clear
distinction between an exponential decay $J_0(t|\x_0) \propto
e^{-Dt\lambda_0^-}$ in the subcritical regime (blue curve) and
power-law decay $J_0(t|\x_0) \propto t^{-2}$ in the critical regime
(green curve).  While we have no theoretical prediction on the
long-time asymptotic behavior of $J_0(t|\x_0)$ in the supercritical
regime, we observe that the two curves for the subcritical and
supercritical regimes are very close to each other.  Further analysis
is needed to rationalize this observation.  We also present the
empirical probability densities estimated directly from Monte Carlo
simulations, which are in excellent agreement with theory (Monte Carlo
results for the supercritical regime are also in agreement with theory
but they are not shown for a better visualization).

\section{Discussion and conclusion}
\label{sec:discussion}

In this paper, we presented a systematic study of surface-mediated
branching processes for ordinary diffusion in bounded domains.  The
competition between absorption and branching events results in a
sophisticated dynamics of the population of particles.  We focused on
the population size $\N(t)$ and its probabilistic characterization via
the generating function $G_s(t|\x_0)$, the probabilities
$Q_k(t|\x_0)$, and the moments $N_k(t|\x_0)$.  We established three
types of closed-form integral equations for these quantities, which
were based on different single-particle propagators: $P^+(\x,t|\x_0)$,
$P^-(\x,t|\x_0)$, and $P(\x,t|\x_0)$.  These propagators differ in how
the catalytic region is treated for a single particle: as absorbing
for $P^+(\x,t|\x_0)$, as neutral for $P(\x,t|\x_0)$, or as catalytic
for $P^-(\x,t|\x_0)$.  In the first two cases, the branching mechanism
on the catalytic region is incorporated via an additional term in the
integral equation.  This flexibility was particularly helpful in the
long-time asymptotic analysis.  While these integral equations played
the main role in our analysis, an equivalent PDE description with a
nonlinear Robin-type boundary condition was also provided.

The nonlinear nature of the branching mechanism presented the major
difficulty for understanding the behavior of the population dynamics.
The remarkable feature of surface-mediated branching processes is that
the mean population size $N_1(t|\x_0)$ still obeys a linear PDE.  In
fact, this quantity is equivalent to the survival probability in the
framework of conventional diffusion-controlled reactions, but with a
{\it negative} effective reactivity on the catalytic region.  On one
hand, the linear character of the problem allows one to employ
standard tools such as spectral expansions or Laplace transform
techniques.  In particular, the long-time behavior of $N_1(t|\x_0)$ is
fully controlled by the principal eigenvalue $\lambda_0^-$ of the
associated Laplace operator: $N_1(t|\x_0) \propto e^{-Dt\lambda_0^-}$.
In this way, we could distinguish the subcritical ($\lambda_0^- > 0$),
critical ($\lambda_0^- = 0$), and supercritical ($\lambda_0^- < 0$)
regimes.  Moreover, further spectral analysis permitted us to identify
the principal Steklov eigenvalue $\mu_0$ as the critical value of the
catalytic rate $\qc$ that distinguishes these three regimes.  Indeed,
if $\qc < \mu_0$, the branching events are not frequent enough to
maintain the population of particles, and the mean population size
decreases exponentially (i.e., $\lambda_0^- > 0$).  In the opposite
case $\qc > \mu_0$, the branching events are too proliferative, and
the population grows exponentially.  Finally, tuning the catalytic
rate $\qc = \mu_0$ allows one to balance branching and absorption
events, yielding a steady-state critical regime for the mean
population size \cite{Grebenkov26}.

While the above classification into three regimes is informative, the
population dynamics is not reducible to the mean value.  We have shown
that the probability $Q_k(t|\x_0)$ of having $k$ particles at time $t$
vanishes as $t\to \infty$ for any fixed $k > 0$.  This conclusion is
intuitively expected in the subcritical regime when the population
progressively disappears at long times.  In particular, $Q_k(t|\x_0)$
decreases exponentially fast, with the same rate $D\lambda_0^-$, as
the mean population size.
In the critical regime, the situation is subtler because the mean
population size does not vanish, whereas higher-order moments
$N_k(t|\x_0)$ grow as $t^{k-1}$.  This is a manifestation of the
critical nature of this regime: while the population $\N(t)$ vanishes
in the overwhelming majority of random realizations, very large
population sizes may be achieved in a small number of random
realizations to ensure the steady-state limit of the mean value and
the growth of higher-order moments.  Accordingly, the probability of
having at least one particle slowly vanishes: $\bar{Q}_0(t|\x_0)
\propto 1/t$.  This situation is quite common in probability.  For
instance, if a random variable takes the values $0$ and $n$ with
probabilities $1-1/n$ and $1/n$, its expectation is equal to $1$ for
any $n$, even though the probability of getting $0$ approaches $1$ as
$n\to \infty$.  We also conjectured a power-law decrease of the
probabilities $Q_k(t|\x_0)$ as $1/t^2$.

The long-time behavior in the supercritical regime may look even more
surprising.  While the mean population side grows exponentially, the
probability $Q_k(t|\x_0)$ still vanishes for any fixed $k > 0$.  At $t
= 0$, there is a single particle so that $Q_k(0|\x_0) = \delta_{k,1}$.
For any $k \geq 2$ fixed, the probability $Q_k(t|\x_0)$ starts
growing, reaches a maximum, and then vanishes, as it is getting more
and more unlikely to keep a fixed number of particles in the growing
population.  Curiously, the probability of having no particles
$Q_0(t|\x_0)$ does not approach $1$ as $t\to \infty$, in contrast to
two other regimes.  We established a nonlinear integral equation for
its limit $Q_0(\infty|\x_0)$.  The limiting value $Q_0(\infty|\x_0) <
1$ may seem to violate the probability normalization.  This is a
consequence of the non-commuting limits $t\to \infty$ and $k\to
\infty$.  In fact, the normalization is preserved for any $t$, but
does not necessarily hold in the limit $t = \infty$.  A similar
``paradoxical'' situation is well known for a Gaussian propagator
$e^{-x^2/(4Dt)}/\sqrt{4\pi Dt}$ for ordinary diffusion on the line:
the propagator is correctly normalized for any $t > 0$, but it
formally vanishes as $t\to \infty$, thus ``violating'' the
normalization at $t=\infty$.

Another specific feature of the supercritical regime is the long-time
behavior of the moments: $N_k(t|\x_0) \propto e^{kDt|\lambda_0^-|}$.
As a consequence, the moments of the rescaled population size,
$\eta(t) = \N(t)/N_1(t|\x_0)$, have finite limits, suggesting that the
stochastic process $\eta(t)$ reaches a steady-state distribution at
long times.  Once this distribution is found, the long-time behavior
of the population dynamics would be fully characterized by the mean
population size.  An exact computation of this limiting distribution
remains an open problem.

The above theoretical conclusions were validated by two independent
numerical techniques: (i) a numerical solution of the integral
equations, and (ii) Monte Carlo simulations.  In order to facilitate
the numerical treatment, we focused on diffusion in a circular
annulus, for which the propagators and the survival probabilities can
be accurately computed via the known spectral expansions or via the
inverse Laplace transform of exact formulas.  The major practical
simplification came from the rotational invariance of the annulus so
that the remaining radial-coordinate problem was one-dimensional, in
which the catalytic and absorbing regions were reduced to two points
at $|\x| = R$ and $|\x| = L$.  This simplification eliminated the
spatial integration in the integral equation that was reduced to a
nonlinear Volterra equation with a weakly singular kernel.  We
proposed a basic scheme for its numerical solution.  This computation
can be easily extended to other basic examples such as an interval or
a concentric spherical shell.  Further improvements of the proposed
numerical technique and its extension to general domains present an
interesting perspective.  In particular, one can employ finite-element
methods for solving nonlinear PDEs.

In this light, Monte Carlo simulations may be considered as a suitable
alternative to finite-element methods.  While simulations of random
absorption and branching events for each particle can be realized by
existing tools (see, e.g., \cite{Schumm23,Ye25}), an accurate
approximation of the population size may be challenging, especially at
long times.  The practical difficulties depend on the considered
regime: (i) in the subcritical regime, the population vanishes so that
an accurate estimation of exponentially small values may be hard due
to the usual statistical error $\propto 1/\sqrt{M}$ for $M$ random
realizations; (ii) in the supercritical regime, the number of
particles grows rapidly thus raising memory-overflow issues and
dramatic slowing down of each simulation; and (iii) in the critical
regime, minor errors in the simulation procedure may switch the
dynamics to subcritical or supercritical regime, and eventual
deviations would be amplified with time.  This brief outline urges for
future developments of accurate numerical techniques for simulating
surface-mediated branching processes.

The presented study can be further extended in various
directions.  (i) The ordinary diffusion can be replaced by a general
Markov process governed by a Fokker-Planck operator
\cite{Grebenkov26a}; the effect of external forces or diffusion
heterogeneities onto the population dynamics can thus be inspected.
(ii) The diffusion equation that governs the time evolution of the
generating function and of all the related quantities may incorporate
an additional term $-\nu G_s(t|\x_0)$ to describe eventual first-order
reactions in the bulk with the rate $\nu$; in this way, one can treat
``mortal'' particles with an exponentially distributed finite lifetime
\cite{Yuste13,Meerson15,Grebenkov17}.  (iii) While we focused on the
stochastic dynamics of the number of particles, spatial aspects of the
autocatalytic processes can be further analyzed, in particular, the
role of proximity to the catalytic region.  (iv) Moreover, temporal
aspects of this dynamics are also accessible such as the
first-extinction time discussed in Sec. \ref{sec:extinction}; in a
similar way, one can investigate the first-crossing times when the
population size exceeds a prescribed threshold.  (v) An extension to
the initial state with many particles presents an interesting
direction.  (vi) Finally, various optimization problems emerge such as
the geometric control of the population growth by reshaping the domain
or redistributing the absorption regions (see \cite{Grebenkov26}).

In summary, surface-mediated branching processes represent an
important and yet poorly understood class of out-of-equilibrium
stochastic dynamics, with potential applications in chemical physics,
heterogeneous catalysis, and population dynamics.  For example, an
extension to diffusion in exterior domains brings open mathematical
problems even for the mean population size.  Furthermore, the
intrinsic out-of-equilibrium nature of these systems offers a
promising perspective for non-equilibrium physics.  The negative
reactivity naturally raises the fundamental question of
time-reversibility at reactive boundaries.  Evaluating the
thermodynamic cost required to sustain such surface-mediated critical
states, and quantifying the dynamic irreversibility driven by the
asymmetry between boundary absorption and branching, remain
fascinating for future research.

\begin{acknowledgments}
D.S.G. thanks Prof. M. Levitin and Prof. A. Hassannezhad for fruitful
discussions.  D.S.G. acknowledges the Simons Foundation for supporting
his sabbatical sojourn in 2024 at the CRM (University of Montr\'eal,
Canada), as well as the Alexander von Humboldt Foundation for support
within a Bessel Prize award.
\end{acknowledgments}

%%%%%%%%%%%%%%%%%%%%%%%%%%%%%%%%%%%%%%
\appendix

\section{Probabilistic description of absorption and branching events}
\label{sec:Amodel}

In this Appendix, we provide a more accurate probabilistic description
of absorption and branching events.  

In the conventional framework of a first-order bulk reaction with a
rate $\nu$, a particle would react in the bulk during a time interval
$(t,t+dt)$ with the probability $\nu dt$.  As we consider here
exclusively surface reactions, the bulk reaction rate $\nu$ should be
replaced by a surface reaction rate $\qa > 0$, whereas the physical
time $t$ is substituted by the boundary local time $\ell_{t,\rm a}$
spent by the particle on the absorbing region $\Gammaa$
\cite{Grebenkov20}.  The latter can be defined as the properly
rescaled residence time of the particle in a thin layer of width $\ve$
near the absorbing region:
\begin{equation}
\ell_{t,\rm a} = \lim\limits_{\ve\to 0} \frac{D}{\ve} \int\limits_0^t dt' \, \Theta(\ve - |\X_{t'} - \Gammaa|),
\end{equation}
where $|\X_{t'} - \Gammaa|$ is the Euclidean distance between the
random position $\X_{t'}$ of the particle at time $t'$ and the
absorbing region $\Gammaa$, and $\Theta(z)$ is the Heaviside step
function: $\Theta(z) = 1$ for $z > 0$ and $0$ otherwise
\cite{Ito,Freidlin}.  In this definition, the boundary local
time $\ell_{t,\rm a}$, despite its name, has units of length;
accordingly, the surface reaction rate $\qa$, despite its name, has
units of inverse length.  In analogy with bulk reactions, at each
arrival onto the absorbing region $\Gammaa$, the particle may
disappear with the probability $\qa d\ell_{t,\rm a}$, so that $\qa$ is
the surface reaction rate per unit boundary local time $\ell_{t,\rm
a}$.
In this Poissonian setting, the absorption event occurs at a random
time 
\begin{equation}
\tau_{\a} = \inf\{ t > 0 ~:~ \ell_{t,\rm a} > \hat{\ell}_{\rm a}\} 
\end{equation}
when the boundary local time $\ell_{t,\rm a}$ exceeds an independent
random threshold $\hat{\ell}_{\rm a}$ with the exponential
distribution $\P\{ \hat{\ell}_{\rm a} > \ell\} = e^{-\qa \ell}$.  This
definition is a direct analogy of the exponential random lifetime
$\hat{t}$ of a particle disappearing in the bulk, $\P\{\hat{t} > t\} =
e^{-\nu t}$, with a given rate $\nu$.  The above probabilistic
description corresponds to a partially reactive boundary (with a
reactivity $\kappa = \qa D$) and yields a standard Robin boundary
condition for macroscopic quantities such as the survival probability
\cite{Grebenkov20}.

In the same vein, branching events on the catalytic region $\Gammac$
can be described by using the boundary local time $\ell_{t,\rm c}$ on
$\Gammac$:
\begin{equation}
\ell_{t,\rm c} = \lim\limits_{\ve\to 0} \frac{D}{\ve} \int\limits_0^t dt' \, \Theta(\ve - |\X_{t'} - \Gammac|).
\end{equation}
At each arrival onto $\Gammac$, the particle may branch into two
identical copies (offsprings) with the surface branching rate $\qc
\geq 0$ per unit boundary local time $\ell_{t,\rm c}$.  In other
words, the branching event occurs with the probability $\qc
d\ell_{t,\rm c}$ at each arrival onto $\Gammac$ or, equivalently, 
at a random time 
\begin{equation}
\tau_{\c} = \inf\{ t > 0 ~:~ \ell_{t,\rm c} > \hat{\ell}_{\rm c}\} 
\end{equation}
when $\ell_{t,\rm c}$ exceeds an independent random threshold
$\hat{\ell}_{\rm c}$ with the exponential distribution: $\P\{
\hat{\ell}_{\rm c} > \ell\} = e^{-\qc \ell}$.  Two newborn particles
are released from the location of the branching event and diffuse
independently.  Each of these particles will disappear or branch at a
later time, and so on.

Even though the above probabilistic constructions of individual
absorption and branching events are identical (except that $\qa$ and
$\qc$ may take different values), their effects onto the population
dynamics are dramatically distinct, as described in the main text.

\section{Monte Carlo simulations}
\label{sec:MC}

In this Appendix, we briefly describe Monte Carlo simulations that we
developed to illustrate our theoretical results.  As both absorption
and branching events are triggered according to the boundary local
time, we adopt the recent escape-from-a-layer (EFL) approach to
simulate efficiently both the position $\X_t$ of each diffusing
particle and its boundary local times $\ell_{t,\rm a}$ and
$\ell_{t,\rm c}$ \cite{Ye25}.  When the particle diffuses away from
the boundary, its boundary local times do not change, and its random
trajectory can be sampled by using the classical walk-on-spheres
algorithm \cite{Muller56}.  In turn, when the particle arrives into a
thin boundary layer, a basic simulation of multiple reflections on the
boundary would require very small timesteps.  We replace this
time-consuming simulation step by a single escape of the particle from
the boundary layer, with an appropriate random increment of the
acquired boundary local time.  This technique was shown to be
efficient for estimating various diffusion-reaction quantities such as
the first-reaction time distribution.

As the absorption and catalytic rates $q_{\a}$ and $q_{\c}$ can be
significantly different, we allow for using different widths
$\ve_{\a}$ and $\ve_{c}$ of boundary layers near the absorbing and
catalytic regions $\Gamma_{\a}$ and $\Gamma_{\c}$.  We use the EFL
approach for simulating diffusion of each particle until its
absorption event on $\Gamma_{\a}$ or branching event on $\Gamma_{\c}$.
The simulation for one particle is stopped when either $\ell_{t,\a}$
exceeds a random threshold $\hat{\ell}_{\a}$ (drawn from the
exponential distribution with the rate $q_{\a}$), or $\ell_{t,\c}$
exceeds another random threshold $\hat{\ell}_{\c}$ (drawn from the
exponential distribution with the rate $q_{\c}$), as defined in
Appendix \ref{sec:Amodel}.  In both cases, the particle is killed, and
one records its position and the time interval from its release time
$t_{\rm ini}$ to the (current) stopping time $t_{\rm end}$.  If the
particle was stopped on the catalytic region, the branching event
occurs, i.e., two new particles are released from the recorded
position, and the above simulation is repeated for each of them.  In
this way, one creates a list of particles with their release and
stopping times.  The simulation is continued until either there is no
active particle (extinction of the population), or the total number of
active particles exceeds a prescribed limit $n_{\rm max}$, or if the
release times of all active particles exceed a prescribed limit
$t_{\rm max}$.  Two last conditions are imposed artificially to handle
an eventual explosive growth of the population size.  Once the
simulation is stopped, one can use the created list to evaluate the
number of particles co-existed at any given time $t$.  Repeating the
above simulation $M$ times, one can estimate the distribution of the
population size $\N(t)$, its moments $N_k(t|\x_0)$, the generating
function $G_s(t|\x_0)$, and the probabilities $Q_k(t|\x_0)$ at any
time $t$.

For validation purposes, we focus on the geometric setting discussed
in the main text, namely, a circular annulus between two concentric
circles of radii $R$ and $L$ that represent the catalytic and
absorption regions, respectively.  For illustrations in the main text,
we used $R = 0.1$, $L = 1$, $q_{\a} = \infty$, $D = 1$, and three
values of $q_{\c}$ to cover three regimes.  The parameters of Monte
Carlo simulations are: $M = 10^5$, $n_{\rm max} = 10^5$, $\ve_{\c} =
10^{-4}$, $\ve_{\a} = 10^{-20}$, and $t_{\rm max} = 10$ (since the
outer circle is considered here to be perfectly absorbing, the
simulation of any particle is stopped upon the first arrival onto this
circle, so that the use of a very small thickness $\ve_{\a}$ does not
slow down simulations).  Monte Carlo results are in excellent
agreement with our theoretical predictions, as discussed in the main
text.

\section{Integral representation of a PDE solution}
\label{sec:technical}

In this Appendix, we recall the standard representation of the
solution of an initial-value problem in terms of the associated
propagator.  Let us consider an auxiliary function $U(t'|\x)$, which
satisfies:
\begin{subequations}  \label{eq:U_problem}
\begin{align}  \label{eq:U_diff}
\partial_{t'} U(t'|\x) - D \Delta U(t'|\x) & = 0 \quad \textrm{in}~\Omega, \\  \label{eq:U_qc}
\partial_n U(t'|\x) \pm \qc U(t'|\x) & = f_{\rm c}(t'|\x) \quad \textrm{on}~ \Gammac, \\  
\partial_n U(t'|\x) + \qa U(t'|\x) & = f_{\rm a}(t'|\x) \quad \textrm{on}~ \Gammaa, \\
\partial_n U(t'|\x) & = 0 \quad \textrm{on}~ \Gammar, \\
U(t=0|\x) &= U_0(\x) ,
\end{align}
\end{subequations}
with some given functions $U_0(\x)$, $f_{\rm c}(t'|\x)$, and $f_{\rm
a}(t'|\x)$, and either choice of the sign in front of $\qc$ in
Eq. (\ref{eq:U_qc}).  Multiplying Eq. (\ref{eq:U_diff}) by
$P^\pm(\x,t-t'|\x_0)$, multiplying the diffusion equation
$-\partial_{t'} P^\pm(\x,t-t'|\x_0) = D \Delta P^\pm(\x,t-t'|\x_0)$ by
$U(t'|\x)$, subtracting these equations, integrating them over
$\x\in\Omega$, using the Green's formula and the boundary conditions,
one gets
\begin{align*}
& \int\limits_\Omega d\x \bigl[P^\pm(\x,t-t'|\x_0) \partial_{t'} U(t'|\x) \\ 
& \qquad + U(t'|\x) \partial_{t'} P^{\pm}(\x,t-t'|\x_0)\bigr] \\
& = D \int\limits_{\Gammac} d\x\, P^{\pm}(\x,t-t'|\x_0) \, f_{\rm c}(t'|\x) \\
& + D \int\limits_{\Gammaa} d\x\, P^{\pm}(\x,t-t'|\x_0) \, f_{\rm a}(t'|\x),
\end{align*}
whereas all other boundary terms cancelled each other.  Integrating
this equation over $t'$ from $0$ to $t$, one has
\begin{align*}
& \int\limits_{\Omega} d\x \biggl[\underbrace{P^{\pm}(\x,0|\x_0)}_{=\delta(\x-\x_0)} U(t|\x) 
- P^{\pm}(\x,t|\x_0) \underbrace{U(0|\x)}_{=U_0(\x)}\biggr] \\
& = D \int\limits_0^t dt' \int\limits_{\Gammac} d\x\, P^{\pm}(\x,t-t'|\x_0) \, f_{\rm c}(t'|\x) \\
& + D \int\limits_0^t dt' \int\limits_{\Gammaa} d\x\, P^{\pm}(\x,t-t'|\x_0) \, f_{\rm a}(t'|\x),
\end{align*}
from which
\begin{align}  \label{eq:U_identity}
U(t|\x_0) & = \int\limits_{\Omega} d\x P^{\pm}(\x,t|\x_0) \, U_0(\x) \\   \nonumber
& + D \int\limits_0^t dt' \int\limits_{\Gammac} d\x\, P^{\pm}(\x,t-t'|\x_0) \, f_{\rm c}(t'|\x) \\  \nonumber
& + D \int\limits_0^t dt' \int\limits_{\Gammaa} d\x\, P^{\pm}(\x,t-t'|\x_0) \, f_{\rm a}(t'|\x).
\end{align}
A similar representation holds for the propagator $P(\x,t|\x_0)$.

The standard representation (\ref{eq:U_identity}) was used several
times in our derivations.
For instance, 

(i) setting $U(t|\x_0) = G_s(t|\x_0)$, $U_0(\x_0) = G_s(0|\x_0) = s$,
$f_{\rm c}(t|\x) = \qc G_s^2(t|\x)$, $f_{\rm a}(t|\x) = \qa$, and
using the propagator $P^+(\x,t-t'|\x_0)$, one realizes that
Eq. (\ref{eq:U_identity}) is identical with the integral equation
(\ref{eq:Gs_decomposition1}), thus implying the PDE description
(\ref{eq:Gs_PDE});

(ii) setting $U(t|\x_0) = \bar{G}_s(t|\x_0)$, $U_0(\x_0) =
\bar{G}_s(0|\x_0) = 1-s$, $f_{\rm c}(t|\x) = - \qc \bar{G}_s^2(t|\x)$,
$f_{\rm a}(t|\x) = 0$, and using the propagator $P^-(\x,t-t'|\x_0)$,
one sees that Eq. (\ref{eq:U_identity}) is identical with the integral
equation (\ref{eq:barGs_int}), thus implying the PDE description
(\ref{eq:barGs}).

\section{Probabilistic derivation of PDE}
\label{sec:proba}

In this Appendix, we sketch an alternative derivation of
Eqs. (\ref{eq:Gs_PDE}), which is independent of the integral equation
(\ref{eq:Gs_integral}), employs the generator of the underlying
stochastic process and relies on the properties of the boundary local
times.  As its rigorous derivation is exceedingly elaborate and thus
beyond the scope of this paper (see, e.g., \cite{Delmas05}), we
restrict our discussion to the main ideas.

(i) As the number of particles cannot change in the bulk, the time
evolution of the generating function $G_s(t|\x_0)$ obeys the same
diffusion equation as the survival probability in the conventional
setting without catalytic branching:
\begin{equation} 
\partial_t G_s(t|\x_0) = D \Delta G_s(t|\x_0) \quad \textrm{in}~ \Omega,
\end{equation}
where $\Delta$ is the Laplace operator acting on $\x_0$.  This is a
standard backward Fokker-Planck (or Kolmogorov) equation that
represents the time evolution due to diffusive displacements governed
by the Laplace operator (the generator of ordinary diffusion).  This
equation is completed by the initial condition
\begin{equation} 
G_s(0|\x_0) = \E_{\x_0}\{ s^1 \} = s ,
\end{equation}
which states that a single particle was released at time $0$ (strictly
speaking, one has to speak about the terminal condition but they are
equivalent for ordinary diffusion).  As absorption and branching
events occur on the boundary, all the nontrivial features appear in
boundary conditions.

(ii) The key point is that each particle, after branching on
$\Gammac$, produces a tree of descendants, which is independent of the
others.  Even through many particles may be present at time $t$, the
boundary conditions for the generating function can be deduced from
the very first branching event.  To get a boundary condition, we set
$\x_0 \in \Gammac$ at time $0$ and consider two options: (1) the
branching event may occur with the probability $\qc d\ell_{0,\rm c}$
(see Appendix \ref{sec:Amodel}), in which case two newborn particles
produce their independent trees of descendants, yielding the total
population size $\N(t) = \N_1(t) + \N_2(t)$ and the contribution
$\E_{\x_0}\{ s^{\N_1(t) + \N_2(t)} \} = [G_s(t|\x_0)]^2$; or (2) the
particle leaves the proximity of $\Gammac$ without branching, keeping
a single tree of descendants for a while (until the next return to
$\Gammac$) and thus contributing $G_s(t|\x_0)$.  As a consequence, the
generating function after this initial branching attempt is
\begin{equation*}
G_s(t|\x_0) \bigl(1 - \qc d\ell_{0,\rm c}\bigr) + [G_s(t|\x_0)]^2\, \qc d\ell_{0,\rm c},
\end{equation*}
i.e., its change before and after an infinitesimal increment
$d\ell_{0,\rm c}$ is
\begin{equation*}
d G_s(t|\x_0) = \qc d\ell_{0,\rm c} \bigl([G_s(t|\x_0)]^2 - G_s(t|\x_0)\bigr).
\end{equation*}
If the right-hand side was proportional to an infinitesimal increment
$dt$ of physical time $t$, one could divide this relation by $dt$ and
take the limit $dt\to 0$ to express the left-hand side as a time
derivative.  In our case, this change occurs on the boundary and is
proportional to the infinitesimal increment of the boundary local time
that results in the {\it spatial} normal derivative:
\begin{equation}  
\partial_n G_s(t|\x_0) = \qc \bigl([G_s(t|\x_0)]^2 - G_s(t|\x_0)\bigr)  \quad \textrm{on}~ \Gammac. 
\end{equation} 
This is a {\it nonlinear} Robin-type boundary condition, which is
reminiscent to the boundary-catalytic branching.  Note that if the
particle was split into $m$ offsprings, the quadratic term would be
replaced by $[G_s(t|\x_0)]^m$.

(iii) A similar argument is applicable to the absorbing region
$\Gammaa$, in which case the absorption event destroys the particle and
thus gives the contribution $s^0 = 1$ instead of $[G_s(t|\x_0)]^2$,
resulting in a linear Robin boundary condition:
\begin{equation}  
\partial_n G_s(t|\x_0) = \qa \bigl[1 - G_s(t|\x_0)\bigr]  \quad \textrm{on}~ \Gammaa. 
\end{equation} 
Finally, no change in $\N(t)$ happens on the reflecting region
$\Gammar$, that yields the Neumann boundary condition:
\begin{equation}  
\partial_n G_s(t|\x_0) = 0  \quad \textrm{on}~ \Gammar. 
\end{equation}

\section{Interpretation of the propagator}
\label{sec:Pminus}

As discussed in Sec. \ref{sec:integral}, the conventional propagator
$P^+(\x,t|\x_0)$ has a standard probabilistic interpretation as the
probability density of finding the particle in a vicinity of point
$\x$ at time $t$, given that it started from $\x_0$ at time $0$ and
has not reacted on $\Gammaa$, nor on $\Gammac$.  In contrast, an
eventual exponential growth of the propagator $P^-(\x,t|\x_0)$ urges
for another interpretation for this quantity.

In the Supplementary Material to \cite{Grebenkov26a}, the formalism of
surface-mediated autocatalytic dynamics was extended to describe the
statistics of the number of particles $\N^A(t)$ in a subset $A\in
\Omega$.  In particular, the mean population size $N^A_1(t|\x_0) =
\E_{\x_0}\{\N^A(t)\}$ was shown to satisfy the initial-value problem:
\begin{subequations}  \label{eq:NA_PDE}
\begin{align}
\partial_t N^A_1 - D \Delta N^A_1 & = 0 \quad \textrm{in} ~ \Omega, \\
\partial_n N^A_1 - \qc N^A_1 & = 0 \quad \textrm{on}~ \Gammac, \\ 
\partial_n N^A_1 + \qa N^A_1 & = 0 \quad \textrm{on}~ \Gammaa, \\
\partial_n N^A_1 & = 0 \quad \textrm{on}~ \Gammar, \\
N^A_1(0|\x_0) & = 1_A(\x_0) ,
\end{align}
\end{subequations}
where $1_A(\x_0)$ is the indicator function of $A$.  Comparison of
this PDE with Eqs. (\ref{eq:Nt_system}) for $N_1(t|\x_0)$ reveals the
only difference in the initial condition.  Moreover, as the PDE
problem (\ref{eq:NA_PDE}) is linear, one can write its solution as
\begin{equation}
N^A_1(t|\x_0) = \int\limits_{A} d\x \, P^-(\x,t|\x_0). 
\end{equation}
As the mean population size in any subset $A$ of the domain can be
obtained by integrating the propagator over $A$, $P^-(\x,t|\x_0)d\x $
can be interpreted as the mean population size in a $d\x$ vicinity of
the point $\x$.  In other words, $P^-(\x,t|\x_0)$ represents the
density (or concentration) of particles at time $t$ for a population
that was initiated by a single particle released from $\x_0$ at time
$0$.  As a consequence, the propagator $P^-(\x,t|\x_0)$ provides the
spatial information on the autocatalytic dynamics.

\section{Circular annulus}
\label{sec:annulus}

As a basic example, we consider the circular annulus $\Omega =
\{\x\in\R^2 ~:~ R < |\x| < L\}$ with the catalytic region $\Gammac$ at
the inner circle of radius $R$ and the absorbing region $\Gammaa$ at
the outer circle of radius $L$ (here $\Gammar = \emptyset$).  In
Secs. \ref{sec:annulus_Steklov} and \ref{sec:annulus_eigen}, we
determine the principal Steklov eigenmode and the principal Laplacian
eigenmode that control the asymptotic behavior.  Section
\ref{sec:propagator} shows the computation of the Laplace-transformed
propagator, whereas Sections \ref{sec:annulus_N1} and
\ref{sec:annulus_Gs_asympt} present the results for the mean
population size and the long-time limit of the generating function,
respectively.

\subsection{Principal Steklov eigenmode}
\label{sec:annulus_Steklov}

As discussed in the main text, the principal eigenvalue $\mu_0$ of the
Steklov problem (\ref{eq:vk_Steklov}) determines the condition for the
critical regime.  As the principal eigenfunction is rotationally
invariant, one can search it as $v_0(\x) = A(\ln(|\x|/R) + B)$, with
two unknown constants $A$ and $B$.  Two boundary conditions fix the
eigenvalue,
\begin{equation}  \label{eq:mu0}
\mu_0 = \frac{1}{R[1/(\qa L) + \ln(L/R)]} \,,
\end{equation}
and the constant $B$ so that
\begin{equation}
v_0(\x) = A \biggl(\ln (L/|\x|) + \frac{1}{\qa L}\biggr).
\end{equation}
The remaining constant $A$ is found from the normalization condition:
\begin{equation*}
1 = \int\limits_{\Gammac} d\x \, v_0^2(\x) = 2\pi R A^2 (1/\mu_0 R)^2
\end{equation*}
and thus
\begin{equation}
A = \mu_0 \sqrt{R/(2\pi)} \,.
\end{equation}
We also evaluate two integrals
\begin{align}  \nonumber
& \int\limits_{\Omega} d\x \, v_0^2(\x) = \mu_0^2 R
\biggl[\frac{L^2-R^2}{8} \bigl((1+2/(\qa L))^2+1\bigr) \\
& \qquad - \frac{R^2\ln(L/R)}{2} \bigl(1 + 2/(\qa L) + \ln(L/R)\bigr) \biggr]
\end{align}
and
\begin{align}  \nonumber
\int\limits_{\Omega} d\x \, v_0(\x) & = \frac{\sqrt{2\pi R} \,\mu_0}{4} 
\biggl[(L^2-R^2)(1+2/(\qa L)) \\
& - 2R^2 \ln(L/R)\biggr].
\end{align}

For a given rate $\qa$, the eigenvalue $\mu_0$ distinguishes the
subcritical ($\qc < \mu_0$), critical ($\qc = \mu_0$), and
supercritical ($\qc > \mu_0$) regimes.  In the limit $\qa = \infty$,
we have
\begin{equation}  \label{eq:qccrit}
\mu_0 = \frac{1}{R\ln(L/R)} = \qccrit 
\end{equation}
and
\begin{equation}
\int\limits_{\Omega} d\x \, v_0^2(\x) = \frac{L^2-R^2 -2R^2\ln(L/R)(1 + \ln(L/R))}{4 R[\ln(L/R)]^2}  \,.
\end{equation}

\subsection{Principal Laplacian eigenmode}
\label{sec:annulus_eigen}

In this section, we summarize the standard formulas to compute the
principal eigenvalue $\lambda_0^-$ and the associated eigenfunction
$u_0^-(\x)$ by solving Eqs. (\ref{eq:Laplace}).  While this
computation can be easily adjusted to get other Laplacian eigenmodes
\cite{Carslaw,Thambynayagam,Grebenkov13}, we focus on the principal
eigenmode that determines the long-time behavior.  Moreover, the
derived formulas are valid for the principal eigenpair $\{\lambda_0^+,
u_0^+(\x)\}$ of Eqs. (\ref{eq:Laplace0}) by replacing $\qc$ by $-\qc$,
and for the principal eigenpair $\{\lambda_0, u_0(\x)\}$ associated to
$\qc = 0$ and satisfying
\begin{subequations}  \label{eq:Laplace00}
\begin{align}
-\Delta u_k & = \lambda_k u_k  \quad \textrm{in}~\Omega, \\
\partial_n u_k  & = 0 \quad \textrm{on}~ \Gammac, \\ 
\partial_n u_k + \qa u_k & = 0 \quad \textrm{on}~ \Gammaa, \\
\partial_n u_k & = 0 \quad \textrm{on}~ \Gammar. 
\end{align}
\end{subequations}
In the following, we treat separately three regimes.

\subsubsection*{Subcritical regime}

In the subcritical regime ($\qc < \mu_0$), the principal eigenvalue
$\lambda_0^-$ is strictly positive.  A rotationally invariant
Laplacian eigenfunction $u^-(\x)$ can be searched as a linear
combination $c_1 J_0(\alpha r) + c_2 Y_0(\alpha r)$, where $r = |\x|$,
$J_n(z)$ and $Y_n(z)$ are the Bessel functions of the first and second
kind, and $c_1$, $c_2$ and $\alpha$ are three parameters to be fixed
via two boundary conditions and normalization.  Using the properties
of Bessel functions \cite{Watson}, one has
\begin{equation}
u_0^-(\x) = A\, w_0(|\x|), 
\end{equation}
where
\begin{align} \nonumber
w_0(r) & = \bigl[-\alpha Y_1(\alpha R) + \qc Y_0(\alpha R)\bigr] J_0(\alpha r) \\
& - \bigl[-\alpha J_1(\alpha R) + \qc J_0(\alpha R)\bigr] Y_0(\alpha r) ,
\end{align}
and $A$ is the normalization constant such that
\begin{equation}
1 = \int\limits_{\Omega} d\x \, [u_0^-(\x)]^2 = A^2 2\pi \int\limits_R^L dr \, r \, w_0^2(r).
\end{equation}
In turn, the principal eigenvalue is determined as $\lambda_0^- =
\alpha^2$, where $\alpha$ is the smallest positive zero of the equation
\begin{align} \nonumber
& \bigl(-\alpha Y_1(\alpha L) + \qa Y_0(\alpha L)\bigr) \bigl(-\alpha J_1(\alpha R) + \qc J_0(\alpha R)\bigr) \\
& = \bigl(-\alpha J_1(\alpha L) + \qa J_0(\alpha L)\bigr) \bigl(-\alpha Y_1(\alpha R) + \qc Y_0(\alpha R)\bigr).
\end{align}
When $\qa = \infty$, it is reduced to
\begin{align} \nonumber
& Y_0(\alpha L) \bigl(-\alpha J_1(\alpha R) + \qc J_0(\alpha R)\bigr) \\
& = J_0(\alpha L) \bigl(-\alpha Y_1(\alpha R) + \qc Y_0(\alpha R)\bigr).
\end{align}

Using the properties of Bessel functions \cite{Watson}, one gets
\begin{align} \nonumber
\int\limits_R^L dr \, r \, w_0^2(r) & = \frac{1}{2\alpha^2} \biggl(R^2 \bigl( [w'_0(R)]^2 + \alpha^2 [w_0(R)]^2 \bigr) \\
& - L^2 \bigl( [w'_0(L)]^2 + \alpha^2 [w_0(L)]^2\bigr) \biggr) . 
\end{align}
We have $w_0(R) = 2/(\pi R)$, $w'_0(R) = - \qc w_0(R)$, and $w'_0(L) =
-\qa w_0(L)$ so that
\begin{align} \nonumber
\int\limits_R^L dr \, r \, w_0^2(r) & = \frac{(\qc^2 + \alpha^2)\frac{4}{\pi^2} - L^2 (\qa^2 + \alpha^2) [w_0(L)]^2}{2\alpha^2} \,. 
\end{align}
For $\qa = \infty$, this relation becomes
\begin{align} \nonumber
\int\limits_R^L dr \, r \, w_0^2(r) & = \frac{(\qc^2 + \alpha^2)\frac{4}{\pi^2} - L^2 [w'_0(L)]^2}{2\alpha^2} \,. 
\end{align}
We also have
\begin{equation}
C_0^- = \int\limits_{\Omega} d\x \, u_0^-(\x) = 2\pi A \frac{R w'_0(R) - L w'_0(L)}{\alpha^2} \,.
\end{equation}

\subsubsection*{Critical regime}

In the critical regime ($\lambda_0^- = 0$), the PDE problems
(\ref{eq:Laplace}) and (\ref{eq:vk_Steklov}) for the Laplacian and
Steklov eigenmodes are identical, so that the principal Laplacian
eigenfunction $u_0^-$ is actually proportional to the principal
Steklov eigenfunction $v_0$ (we recall that these eigenfunctions have
different normalizations):
\begin{equation}
u_0^-(\x) = A \biggl(\ln(L/|\x|) + \frac{1}{\qa L}\biggr),
\end{equation}
where the normalization constant $A$ is obtained from 
\begin{align*}
1 & = \int\limits_\Omega d\x \, [u_0^-(\x)]^2 = A^2 2\pi \int\limits_R^L dr \, r \, \bigl(\ln(L/r) + 1/(\qa L)\bigr)^2 \\
& = \frac{\pi A^2}{4} \biggl(L^2 \bigl[(2/(\qa L)+1)^2 + 1\bigr] \\
& - R^2 \bigl[(2/(\qa L)+1 + 2\ln(L/R))^2+1\bigr]\biggr).
\end{align*}
Integrating $u_0^-(\x)$ over $\Omega$, we also get
\begin{equation}
C_0^- = \pi A \biggl((L^2-R^2)(1/2 + 1/(\qa L)) - R^2\ln(L/R) \biggr).
\end{equation} 
In the limit $\qa = \infty$, we have
\begin{align*}
A^{-2} = \frac{\pi}{2} \biggl(L^2 - R^2 - 2R^2 \ln(L/R)(1 + \ln(L/R))\biggr)
\end{align*}
and
\begin{equation}
C_0^- = \frac{\pi}{2} A \bigl(L^2-R^2 - 2R^2\ln(L/R) \bigr).
\end{equation}

\subsubsection*{Supercritical regime}

In the supercritical regime ($\qc > \mu_0$), the principal eigenvalue
$\lambda_0^-$ is negative, so that $J_n(z)$ and $Y_n(z)$ should be
replaced by modified Bessel functions $I_n(z)$ and $K_n(z)$ of the
first and second kind, respectively.  The remaining computation is
almost identical to that in the subcritical regime.  The eigenfunction
reads
\begin{equation}
u_0^-(\x) = A \, w_0(|\x|), 
\end{equation}
where
\begin{align} \nonumber
w_0(r) & = \bigl[-\alpha K_1(\alpha R) + \qc K_0(\alpha R)\bigr] I_0(\alpha r) \\
& - \bigl[\alpha I_1(\alpha R) + \qc I_0(\alpha R)\bigr] K_0(\alpha r) ,
\end{align}
and $A$ is the normalization constant such that
\begin{equation}
1 = \int\limits_{\Omega} d\x \, [u_0^-(\x)]^2 = A^2 2\pi \int\limits_R^L dr \, r \, w_0^2(r).
\end{equation}
The associated eigenvalue is $\lambda_0^- = -\alpha^2$, where $\alpha$
is the smallest positive zero of the equation
\begin{align*}
& \bigl(-\alpha K_1(\alpha L) + \qa K_0(\alpha L)\bigr) \bigl(\alpha I_1(\alpha R) + \qc I_0(\alpha R)\bigr) \\
& = \bigl(\alpha I_1(\alpha L) + \qa I_0(\alpha L)\bigr) \bigl(-\alpha K_1(\alpha R) + \qc K_0(\alpha R)\bigr).
\end{align*}
When $\qa = \infty$, it is reduced to
\begin{align*}
& K_0(\alpha L) \bigl(\alpha I_1(\alpha R) + \qc I_0(\alpha R)\bigr) \\
& = I_0(\alpha L) \bigl(-\alpha K_1(\alpha R) + \qc K_0(\alpha R)\bigr).
\end{align*}

Using the properties of the modified Bessel functions \cite{Watson},
one gets
\begin{align} \nonumber
\int\limits_R^L dr \, r \, w_0^2(r) & = \frac{1}{2\alpha^2} \biggl(R^2 \bigl( [w'_0(R)]^2 - \alpha^2 [w_0(R)]^2 \bigr) \\
& - L^2 \bigl( [w'_0(L)]^2 - \alpha^2 [w_0(L)]^2\bigr) \biggr) . 
\end{align}
We have $w_0(R) = -1/R$, $w'_0(R) = - \qc w_0(R)$, and $w'_0(L) = -\qa
w_0(L)$ so that
\begin{align} \nonumber
\int\limits_R^L dr \, r \, w_0^2(r) & = \frac{(\qc^2 - \alpha^2) - L^2 (\qa^2 - \alpha^2) [w_0(L)]^2}{2\alpha^2} \,. 
\end{align}
For $\qa = \infty$, this relation becomes
\begin{equation} 
\int\limits_R^L dr \, r \, w_0^2(r) = \frac{(\qc^2 - \alpha^2) - L^2 [w'_0(L)]^2}{2\alpha^2} \,. 
\end{equation}
We also have
\begin{equation}
C_0^- = 2\pi A \frac{L w'_0(L) - R w'_0(R)}{\alpha^2} \,.
\end{equation}

\subsection{The propagator}
\label{sec:propagator}

Since the Laplacian eigenmodes are known for the circular annulus, the
propagator $P^-(\x,t|\x_0)$ can be found through the spectral
expansion (\ref{eq:p_spectral}).  To avoid technical details, we focus
on the Laplace-transformed propagator, for which the computation is
simpler:
\begin{equation}
\tilde{P}^-(\x,p|\x_0) = \int\limits_0^\infty dt \, e^{-pt} \, P^-(\x,t|\x_0).
\end{equation}
Moreover, as our setting is rotationally invariant, we average
$\tilde{P}^-(\x,p|\x_0)$ over the angular coordinates.  Its
rotationally-invariant contribution reads (see,
e.g. \cite{Grebenkov21g})
\begin{equation}
\tilde{P}^-(r,p|r_0) = \frac{1}{2\pi D W} \begin{cases} v^L(r_0) \, v^R(r) \quad (r \leq r_0), \cr  
v^L(r) \, v^R(r_0) \quad (r > r_0), \end{cases}
\end{equation}
where $\alpha = \sqrt{p/D}$ and
\begin{subequations}  \label{eq:vR_vL}
\begin{align}  \nonumber
v^R(r) & = (\alpha I_1(\alpha R) + \qc I_0(\alpha R)) K_0(\alpha r) \\
& - (-\alpha K_1(\alpha R) + \qc K_0(\alpha R)) I_0(\alpha r) ,\\  \nonumber
v^L(r) & = (\alpha I_1(\alpha L) + \qa I_0(\alpha L)) K_0(\alpha r) \\
& - (-\alpha K_1(\alpha L) + \qa K_0(\alpha L)) I_0(\alpha r) ,\\  \nonumber
W & = (-\alpha K_1(\alpha L) + \qa K_0(\alpha L)) (\alpha I_1(\alpha R) + \qc I_0(\alpha R)) \\  \label{eq:C_annulus}
& - (-\alpha K_1(\alpha R) + \qc K_0(\alpha R)) (\alpha I_1(\alpha L) + \qa I_0(\alpha L)). 
\end{align}
\end{subequations}
Since $v^R(R) = 1/R$ due to the Wronskian of the modified Bessel
functions, one gets
\begin{equation}  \label{eq:Ptilde_minus}
\tilde{P}^-(R,p|r_0) = \frac{v^L(r_0)}{2\pi R D W} \,.
\end{equation}

Using the properties of modified Bessel functions, we also compute
\begin{align} \nonumber
& \int\limits_R^L dr \, r \, v^L(r) = - \frac{\qa}{\alpha^2} \\ \nonumber
& + \frac{R}{\alpha} \biggl[\alpha \bigl(I_1(\alpha L) K_1(\alpha R) - K_1(\alpha L) I_1(\alpha R)\bigr) \\  \label{eq:vL_int}
& + \qa \bigl(I_0(\alpha L) K_1(\alpha R) + K_0(\alpha L) I_1(\alpha R)\bigr)\biggr]
\end{align}
and
\begin{align} \nonumber
& \int\limits_R^L dr \, r \, v^R(r) = \frac{\qc}{\alpha^2} \\ \nonumber
& + \frac{L}{\alpha}
\biggl[\alpha \bigl(I_1(\alpha L) K_1(\alpha R) - K_1(\alpha L) I_1(\alpha R)\bigr) \\  \label{eq:vR_int}
& - \qc \bigl(I_0(\alpha R) K_1(\alpha L) + K_0(\alpha R) I_1(\alpha L)\bigr)\biggr].
\end{align}

In the limit $\qa = \infty$, $v^R(r)$ remains unchanged, whereas
$v^L(r)$ and $W$ are reduced to
\begin{subequations}  \label{eq:vR_vL_qainf}
\begin{align}  
v^L(r) & = I_0(\alpha L) K_0(\alpha r) - K_0(\alpha L) I_0(\alpha r) ,\\  \nonumber
W & =  K_0(\alpha L) (\alpha I_1(\alpha R) + \qc I_0(\alpha R)) \\  \label{eq:Cinfty}
& - I_0(\alpha L)(-\alpha K_1(\alpha R) + \qc K_0(\alpha R)) . 
\end{align}
\end{subequations}

\subsection{Mean population size}
\label{sec:annulus_N1}

For the circular annulus, an explicit computation of the mean
population size $N_1(t|\x_0)$ is possible in the Laplace domain.  For
this purpose, it is sufficient to integrate the Laplace-transformed
propagator $\tilde{P}^-(\x,p|\x_0)$ over $\Omega$ that yields
\begin{equation}  \label{eq:N1_tilde}
\tilde{N}_1(p|r_0) = \frac{1}{p} \biggl(1 + \frac{\qc \, v^L(r_0) - \qa \, v^R(r_0)}{W}\biggr),
\end{equation}
where $v^L(r_0)$, $v^R(r_0)$ and $W$ are given by
Eqs. (\ref{eq:vR_vL}).  When the starting point is uniformly
distributed in the bulk, one deals with
\begin{align} \label{eq:Np1_annulus}
& \tilde{N}_1(p|\circ) = \frac{2}{L^2-R^2} \int\limits_R^L dr \, r \, \tilde{N}_1(p|r) \\  \nonumber
& = \frac{1}{p} + \frac{2}{(L^2-R^2)pW} \int\limits_R^L dr \, r \bigl[\qc \, v^L(r_0) - \qa \, v^R(r_0)\bigr],
\end{align}
which can be found explicitly by using Eqs. (\ref{eq:vL_int},
\ref{eq:vR_int}).  While it is possible to invert the Laplace
transform by identifying the poles of $\tilde{N}_1(p|r_0)$ and
applying the residue theorem, we employ the numerical inversion via
the Talbot algorithm \cite{Talbot79}.

\subsection{Long-time limit of the generating function}
\label{sec:annulus_Gs_asympt}

Since the annulus is rotationally invariant, the dependence of the
generating function $G_s(t|\x_0)$ on $\x_0$ is reduced to the radial
coordinate $r_0 = |\x_0|$.  As a consequence, the Green's function
$\G^+(\x,\x_0)$ in the integral equation (\ref{eq:Gs_infty}) is
averaged over the angular coordinate, yielding
\begin{align}  \nonumber
\g(r_0) & = \qc \int\limits_{\Gammac} d\x \, \G^+(\x,\x_0) \\
& = \frac{\ln(L/r_0) + 1/(\qa L)}{\ln(L/R) + 1/(\qc R) + 1/(\qa L)} \,.
\end{align}
Averaging the integral equation (\ref{eq:Gs_infty}) over the angular
coordinate of $\x_0$, we reduce this integral equation to the
algebraic equation:
\begin{equation}  \label{eq:Gs_infty_annulus}
\bar{G}_s(\infty|r_0) = \g(r_0) \,  \bigl(\bar{G}_s(\infty|R) - [\bar{G}_s(\infty|R)]^2\bigr).
\end{equation} 
Setting $r_0 = R$, we arrive at the quadratic equation, which has two
solutions: $\bar{G}_s(\infty|R) = 0$ and
\begin{equation}  \label{eq:barGsR_inf_circular}
\bar{G}_s(\infty|R) = 2 - \frac{1}{\g(R)} = 1 - \frac{1/(\qc R)}{\ln(L/R) + 1/(\qa L)} \,. % = 1 - \frac{\mu_0}{\qc}\,,
\end{equation}
Using Eq. (\ref{eq:mu0}) for the principal Steklov eigenvalue $\mu_0$,
one can rewrite the nontrivial solution as
\begin{equation}
G_s(\infty|R) = \frac{\mu_0}{\qc} \,.
\end{equation}
Since the generating function cannot exceed $1$, the nontrivial
solution is possible if and only if $\qc > \mu_0$, i.e., in the
supercritical regime, in agreement with our general conclusion in
Sec. \ref{sec:Gs_long_growth}.  In turn, $G_s(\infty|R) = 1$ is the
only possible solution for the subcritical and critical regimes.  In
these regimes, Eq. (\ref{eq:Gs_infty_annulus}) immediately implies
that $G_s(\infty|r_0) = 1$ for any starting point.

When the starting point is uniformly distributed, one has
\begin{equation}  \label{eq:Gs_infty_annulus_uni}
G_s(\infty|\circ) = 1 + \g(\circ) \,  \bigl([G_s(\infty|R)]^2 - 1\bigr),
\end{equation} 
where
\begin{equation}
\g(\circ) = \frac{(L^2-R^2)(2/(\qa L)+1) - 2R^2\ln(L/R)}{2(L^2-R^2)[\ln(L/R) + 1/(\qc R) + 1/(\qa L)]} \,.
\end{equation}

We also note that the above analysis was done for $s < 1$.  In turn,
one has $G_1(t|\x_0) = 1$ due to the normalization of probabilities in
any regime.

\section{Numerical solution of integral equations}
\label{sec:numerics}

In this Appendix, we present a numerical scheme for solving integral
equations (\ref{eq:Qn_P0}, \ref{eq:barQ0_P0}) to determine the
probabilities $Q_k(t|\x_0)$, as well as the generating function
$G_s(t|\x_0)$, in the case of the circular annulus with radii $R$ and
$L$.  As this domain is rotationally invariant, $Q_k(t|\x_0)$ and
$G_s(t|\x_0)$ are functions of the radial coordinate $r_0 = |\x_0|$.
As a consequence, the integral equation (\ref{eq:barQ0_P0}) becomes
\begin{align} \nonumber
\bar{Q}_0(t) & = S(t|R) + \qc D 2\pi R \int\limits_0^t dt' \, P(R,t'|R) \\ \label{eq:Q0_P0_bis}
& \times \biggl(\bar{Q}_0(t-t') - \bar{Q}_0^2(t-t')\biggr),
\end{align}
where we used the shortcut notation $\bar{Q}_0(t) = \bar{Q}_0(t|\x_0)$
for any $\x_0 \in \Gammac$ such that $|\x_0| = R$, whereas $S(t|R)$
and $P(R,t'|R)$ are shortcut notations for $S(t|\x_0)$ and
$P(\x,t'|\x_0)$ evaluated at $|\x| = |\x_0| = R$, see
Eqs. (\ref{eq:propagator0}, \ref{eq:S0}).  As the equation for
$\bar{G}_s(t|\x_0)$ just includes the additional factor $(1-s)$ in
front of $S(t|R)$, see Eq. (\ref{eq:barGs_P0}), we do not discuss it
separately.

The above equation (\ref{eq:Q0_P0_bis}) is a nonlinear Volterra
equation of the second kind.  Despite the convolution form, the common
Laplace transform is not efficient because of nonlinearity.  Another
difficulty is that the kernel $P(R,t'|R)$ is weakly singular that
follows from the short-time behavior of the propagator (see below).
In the following, we describe a basic scheme to overcome these issues
and to get a numerical solution of Eq. (\ref{eq:Q0_P0_bis}) and
related integral equations.  We expect that the computational
efficiency of our numerical method may be further improved.

\subsection{Short-time behavior of the kernel}

Let us first discuss the weak singularity of the kernel $P(R,t'|R)$.  To
see this point, we inspect the auxiliary function
\begin{equation}
f(t) = \int\limits_{\Gammac} d\x_0 \int\limits_{\Gammac} d\x \, P(\x,t|\x_0)
\end{equation}
in the general setting.  We assume that $\Gammaa$ and $\Gammac$ are
disconnected regions of the boundary (as in the case of the circular
annulus).  As a consequence, $\Gammaa$ and $\Gammac$ are separated
from each other by a strictly positive distance, so that the
short-time behavior of $f(t)$ is not affected by $\Gammaa$.  The
asymptotics of $f(t)$ as $t\to 0$ can thus be found by considering
reflected Brownian motion started on $\Gammac$.  In the leading order,
the curvature of $\Gammac$ does not matter so that one can approximate
the propagator by that in the upper half-space $\R^d_+$, $P(\x,t|\x_0)
\approx 2 e^{-|\y-\y_0|^2/(4Dt)}/(4\pi Dt)^{d/2}$, where the points
$\x = (\y,0)$ and $\x_0 = (\y_0,0)$ lie on the hyperplane $\partial
\R^d_+$ (the factor $2$ comes from the reflecting condition).  At
short times, the integral over the lateral coordinate $\y$ yields
$1/\sqrt{\pi Dt}$, whereas the second integral over $\y_0$ gives the
area of $\Gammac$:
\begin{equation}  \label{eq:fasympt}
f(t) \simeq \frac{|\Gammac|}{\sqrt{\pi Dt}} + O(1)  \quad (t\to 0).
\end{equation}

In the particular case of the circular annulus, the rotational
invariance implies $f(t) = |\Gammac|^2 P(R,t|R)$, from which
\begin{equation}
P(R,t|R) = \frac{f(t)}{|\Gammac|^2} \simeq \frac{1}{2\pi R \sqrt{\pi Dt}} \quad (t\to 0). 
\end{equation}

In order to treat this weak singularity explicitly, we define
\begin{equation}
p(t) = 2\pi R  \qc D \, P(R,t|R)\, \sqrt{t},
\end{equation}
which has a finite limit as $t\to 0$: $p(0) = \qc \sqrt{D/\pi}$.  Note
also that $S(t|R)$ is exponentially close to $1$ at short times.
We rewrite the above equation (\ref{eq:Q0_P0_bis}) as
\begin{equation}   \label{eq:Q0_P0_tre}
\bar{Q}_0(t) = S(t|R) + \int\limits_0^t dt' \, t'^{-1/2} \, p(t') \, v(t-t'),
\end{equation}
with the initial condition $\bar{Q}_0(0) = 1$, and 
\begin{equation}
v(t) = \bar{Q}_0(t) - \bar{Q}_0^2(t).
\end{equation}

\subsection{Numerical computation of the kernel}

To proceed, we need to evaluate $S(t|R)$ and $p(t)$.  These quantities
can be found either in terms of spectral expansions over the Laplacian
eigenfunctions, or via the inverse Laplace transform.
In the Laplace domain, we can use Eq. (\ref{eq:Ptilde_minus}) to get
$\tilde{P}(R,p|R)$ and Eq. (\ref{eq:N1_tilde}) to get
$\tilde{S}(p|R)$, by setting $\qc = 0$.  We employ the Talbot
algorithm for the numerical inversion of the Laplace transform.  As
both $P(R,t|R)$ and $S(t|R)$ rapidly decay at infinity, one needs a
more accurate computation for large $t$.  In the long-time regime, we
approximate these functions by the leading term of their spectral
expansions:
\begin{align}
P(R,t|R) & \approx [u_0(R)]^2 \,e^{-Dt\lambda_0} , \\
S(t|R) & \approx u_0(R) \, C_0 \, e^{-Dt\lambda_0} , 
\end{align}
where the eigenpair $\{\lambda_0, u_0\}$ was defined in
Eq. (\ref{eq:Laplace00}), $C_0$ is the integral of $u_0$ over
$\Omega$, and their computation is described in
Sec. \ref{sec:annulus_eigen} (where we have to set $\qc = 0$).

\subsection{Discretization}

Choosing a small timestep $\delta = t/n$ and equally spaced grid, we
discretize Eq. (\ref{eq:Q0_P0_tre}) as
\begin{align}
\bar{Q}_0(n\delta) \approx S(n\delta|R) + \sum\limits_{j=0}^n w_j \, v((n-j)\delta) ,
\end{align}
where $w_j$ are suitable weights.  

In a basic quadrature, we use piecewise-constant approximations for
functions $p(t')$ and $v(t-t')$ on each interval $(\delta
j,\delta(j+1))$, so that
\begin{align*}
\int\limits_{t_j}^{t_{j+1}} \frac{dt'}{\sqrt{t'}}\, p(t')\, v(t-t') & \approx \frac{v_{n-j} + v_{n-j-1}}{2} \\
& \times \underbrace{(\sqrt{t_{j+1}} - \sqrt{t_j}) (p_j + p_{j+1}) }_{=a_j} ,
\end{align*}
for each $j\in \{0,1,\ldots,n-1\}$, with $p_j = p(j\delta)$.  As a
consequence, we have
\begin{equation}  \label{eq:wj}
w_j = \frac{a_j + a_{j-1}}{2}  \quad (j=0,1,\ldots,n),
\end{equation}
where we set $a_{-1} = a_n = 0$.

Next, we separate the term with $j = 0$ to get the quadratic equation:
\begin{equation}
\bar{Q}_0(n\delta) = w_0 \bigl(\bar{Q}(n\delta) - \bar{Q}_0^2(n\delta)\bigr) + f_n, 
\end{equation}
where
\begin{equation}
f_n = S(n\delta|R) + \sum\limits_{j=1}^n w_j\, \bigl[\bar{Q}((n-j)\delta) - \bar{Q}_0^2((n-j)\delta)\bigr] .
\end{equation}
A suitable solution of the quadratic equation reads
\begin{equation}
\bar{Q}_0(n\delta) = \frac{2f_n}{1-w_0 + \sqrt{(1-w_0)^2+4w_0 f_n}} \,.
\end{equation}

For the examples shown in the main text, we used $\delta = 10^{-3}$
that required the evaluation of $\bar{Q}_0(n\delta)$ for $n$ ranging
from $1$ to $10000$ to get the probability $\bar{Q}_0(t)$ up to $t =
10$.  The above scheme was implemented in Matlab and took less than a
minute on a personal laptop.  Our numerical results remained almost
unaffected upon diminishing $\delta$ by factor $2$.

\subsection{Solutions for $k > 0$}

In the same way, one can discretize Eq. (\ref{eq:Qn_P0}) with $k > 0$
as
\begin{equation}
Q_k(n\delta) \approx S(n\delta|R) \delta_{k,1} + \sum\limits_{j=0}^n w_j \, v((n-j)\delta)  ,
\end{equation}
with 
\begin{subequations}
\begin{align}
v(t) & = Q_k(t) \bigl(2Q_0(t) - 1\bigr) + v_0(t) , \\
v_0(t) & = \sum\limits_{i=1}^{n-1} Q_i(t) Q_{n-i}(t),
\end{align}
\end{subequations}
where we wrote separately the term $2Q_0(t)Q_k(t)$ from $H_k(t|\x)$.
We get then
\begin{align}
Q_k(n\delta) \approx w_0 Q_k(n\delta) \bigl(2Q_0(n \delta) - 1\bigr) + f_n ,
\end{align}
where
\begin{equation}
f_n = S(n\delta|R) \delta_{k,1} + w_0 v_0(n\delta) + \sum\limits_{j=1}^n w_j \, v((n-j)\delta).
\end{equation}
As a consequence, one obtains
\begin{equation}
Q_k(n\delta) \approx \frac{f_n}{1 - w_0 \bigl(2Q_0(n\delta) - 1\bigr)}  \,.
\end{equation}

\subsection{Computation of the moments}

The same technique can be used to compute the moments $N_k(t|\x_0)$ by
solving the integral equation (\ref{eq:Nk_fk}).  In contrast to the
generating function, the moments $N_k(t|\x_0)$, as well as the kernel
$P^-(\x,t|\x_0)$ in Eq. (\ref{eq:Nk_fk}) grow exponentially in the
supercritical regime.  This issue can be easily resolved by rescaling
the moment $N_k(t|\x_0)$ by an appropriate exponential factor given by
the asymptotic relation (\ref{eq:Ntk_asympt_growth}).  In fact,
setting
\begin{equation}
\bar{N}_k(t|\x_0) = N_k(t|\x_0) e^{-k Dt|\lambda_0^-|} ,
\end{equation}
one can rewrite Eq. (\ref{eq:Nk_fk}) as
\begin{align}  \label{eq:Nk_fk2}
\bar{N}_k(t|\x_0) & = \bar{S}^-(t|\x_0)   \\  \nonumber
& + \qc D \int\limits_{\Gammac} d\x \int\limits_0^t dt' \, \bar{P}^-(\x,t'|\x_0) \, \bar{F}_k(t-t'|\x),
\end{align}
where
\begin{align*}
\bar{S}^-(t|\x_0) & = S^-(t|\x_0) e^{-k Dt|\lambda_0^-|} , \\
\bar{P}^-(\x,t|\x_0) & = P^-(\x,t|\x_0)  e^{-k Dt|\lambda_0^-|}
\end{align*}
and
\begin{align} \nonumber
\bar{F}_k(t|\x) & = F_k(t|\x) e^{-k Dt|\lambda_0^-|}  \\
& = \sum\limits_{j=1}^{k-1} \binom{k}{j} \bar{N}_j(t|\x) \bar{N}_{k-j}(t|\x).
\end{align}
In this way, the integral equation (\ref{eq:Nk_fk2}) does not contain
any exponentially growing functions in the supercritical regime.

Since the right-hand side of Eq. (\ref{eq:Nk_fk2}) does not contain
$\bar{N}_k$, this equation can be solved iteratively, starting from
$\bar{N}_1(t|\x_0)$, which can be obtained directly via the inverse
Laplace transform of the exact expression (\ref{eq:N1_tilde}).
Moreover, one can apply a fast Fourier transform for rapidly
evaluating the convolution in time.  Despite the numerical efficiency
of this technique, we keep using the basic discretization scheme
discussed above that handles the weak singularity of the kernel.

For any fixed $k = 2,3,\ldots$, we define the kernel 
\begin{equation}  \label{eq:pt_Pminus}
p(t) = 2\pi R \qc D \, P^-(R,t|R) e^{-kDt|\lambda_0^-|}\, \sqrt{t}
\end{equation}
and approximate the above integral equation (\ref{eq:Nk_fk2}) as
\begin{equation}
\bar{N}_k(n\delta) \approx \bar{S}^{-}(n\delta|R) + \sum\limits_{j=0}^n w_j \bar{F}_k((n-j)\delta),
\end{equation}
with the weights $w_j$ given by Eq. (\ref{eq:wj}), except that $p_j =
p(\delta j)$ are now determined by the propagator $\bar{P}^-(R,t|R)$
via Eq. (\ref{eq:pt_Pminus}).  Note that the leading-order term in the
short-time behavior of this propagator does not depend on the
reactivity so that $p(0) = \qc \sqrt{D/\pi}$ remains unchanged.

\end{document}